\documentclass[graybox, secnum]{svmult}

\usepackage{amsmath}
\usepackage{amssymb}
\usepackage{mathptmx}       
\usepackage{helvet}         
\usepackage{courier}        
\usepackage{type1cm}        
\usepackage{makeidx}         
\usepackage{graphicx}        
\usepackage{multicol}        
\usepackage[bottom]{footmisc}
\usepackage{hyperref}        
\usepackage{soul}            
\usepackage{color}
\hypersetup{colorlinks=true,urlcolor=blue}
\usepackage[square,numbers]{natbib}
\makeindex             

\begin{document}
\title*{The Merger Dynamics of the X-ray Emitting Plasma in Clusters of Galaxies}
\author{John ZuHone \thanks{corresponding author} and Yuanyuan Su}
\institute{John ZuHone \at Center for Astrophysics $\vert$ Harvard \& Smithsonian, 60 Garden St., Cambridge, MA 02138, \email{john.zuhone@cfa.harvard.edu}
\and Yuanyuan Su \at University of Kentucky, 505 Rose St., Lexington, KY 40506, \email{ysu262@g.uky.edu}}
%
%
\maketitle
\abstract{As the formation of cosmic structure continues to proceed, we observe one of the latest stages of this process in mergers of clusters of galaxies with other clusters, groups, and galaxies. The X-ray emitting hot plasma of these systems can be dramatically affected by these mergers, producing cold and shock fronts, gas sloshing, bulk motions, and turbulence. Combined with numerical simulations, observations of these features can be used to constrain the plasma physics of the hot gas as well as its interaction with high-energy cosmic rays. In this chapter, we review these topics and point forward to what the capabilities of future observatories will reveal.}


\section{Introduction}

The growth of structure on cosmological scales due to gravity has occurred in a ``bottom-up'' fashion, with less massive objects merging to become larger objects. \textit{Mergers} between clusters of galaxies are the latest stage of this process and the most energetic events in the current universe. They involve the conversion of up to $\sim$10$^{65}$~erg of gravitational potential energy into the kinetic energy of dark matter (DM), the hot \textit{intracluster medium (ICM)}, and galaxies, as well as into the thermal and magnetic energy of the ICM and the acceleration of \textit{cosmic ray (CR)} particles. 

If clusters did not merge with each other, they would be relatively passive objects, aside from accretion of material at their outskirts and AGN feedback at their centers. During a cluster merger, gravitational potential energy is converted into kinetic, thermal, magnetic and CR energy to produce a variety of features. These include X-ray observables such as cold fronts (Section \ref{sec:cold_fronts}), 
shock fronts (Section \ref{sec:shock_fronts}),
and stripped gas tails (Section \ref{sec:tails}),  all of which can be used to probe ICM properties.

There are also important synergies with radio observations to be exploited in merging clusters. High-energy CR electrons emit radio synchrotron emission in the ICM magnetic field, and in merging clusters these particles produce features such as bent radio lobes, radio halos/mini-halos, and radio relics that are tightly linked with X-ray properties because these CRe interact with the ICM plasma.

In this chapter, we will discuss the properties of cluster mergers from the perspective of X-ray observations, along with the advances which have been made by examining these observations in the light of theory and simulations. 

\section{X-ray Features Produced By Cluster Mergers}

Thanks to the latest generation of X-ray telescopes, a number of tell-tale features in the ICM of merging clusters have been identified, especially connected with the surface brightness and temperature distributions of the cluster gas. These features are confirmed in numerical simulations of cluster mergers, providing an opportunity to use them to constrain the physics of the ICM. 

\subsection{Cold Fronts}\label{sec:cold_fronts}

\textit{Cold fronts} are some of the most prominent features in the ICM due to their brightness and prevalence in many galaxy clusters. Cold fronts are sharp \textit{contact discontinuities}, where the denser side of the front is colder. Unlike a \textit{shock front} (Section \ref{sec:shock_fronts}) with a large pressure jump, there is no significant pressure change across a cold front. In X-ray observations, they are characterized by a surface brightness edge separating the hot and diffuse ICM and the relatively cold and dense gas within the substructure. These were first discovered by \textit{Chandra} thanks to its sub-arcsecond angular resolution. Most cold fronts are formed as a consequence of merging activity, whether major or minor. Because of the frequency of mergers, most observed clusters have at least one cold front \cite{Ghizzardi2010}. 

Cold fronts can potentially provide important information on the microscopic conditions of the ICM. Neglecting for the moment the effects of magnetic fields, particle interactions in a hot, ionized plasma such as the ICM are mediated by the electric force. The so-called ``\textit{Coulomb collisions}'' between electron-electron, ion-ion, or electron-ion pairs set a length scale below which we expect the fluid approximation to break down and for gradients in the fluid properties to be smeared out by particle diffusion. This length scale is the ``\textit{mean free path}'', or average length between collisions \cite{Spitzer1962}:
\begin{equation}
\lambda_{\rm mfp} = \frac{3^{3/2}(k_BT)^2}{4\pi^{1/2}nq^4\ln{\Lambda}} \sim 1-10~\rm{kpc}
\end{equation}
where $n$ is the electron/ion number density, and $q$ is the electron/ion charge. $\ln\Lambda$ is the ``Coulomb logarithm'', where $\Lambda$ is the ratio of largest to smallest impact parameters for particle collisions, and has typical values of $\sim$30-40. As we shall see, the widths of most cold front interfaces are smaller than this, indicating some process is preventing particles to diffuse across the interface and reduce the sharpness of the front jump.

Similar considerations lead to the expected values for the \textit{viscosity} and \textit{thermal conductivity} of the plasma  \cite{Spitzer1962}:
\begin{eqnarray}
\eta &\approx& 0.58n_i\lambda_im_i\left(\frac{k_BT}{m_i}\right)^{1/2}~~~{\rm (viscosity)} \label{eqn:visc}\\ 
\nonumber &\approx& 5500~{\rm g}~{\rm s}^{-1}~{\rm cm}^{-1}\left(\frac{T}{10^8~K}\right)^{5/2}\left(\frac{\ln{\Lambda}}{40}\right)^{-1} \\
\kappa &\approx& 1.31n_e\lambda_ek_B\left(
\frac{k_BT}{m_e}\right)^{1/2}~~~{\rm (thermal~conductivity)}  \label{eqn:cond}\\
\nonumber &\approx& 4.6 \times 10^{13}~{\rm erg}~{\rm s}^{-1}~{\rm cm}^{-1}~{\rm K}^{-1}\left(\frac{T}{10^8~K}\right)^{5/2}\left(\frac{\ln{\Lambda}}{40}\right)^{-1} 
\end{eqnarray}
where $m_{i/e}$ is the mass of the ion/electron, $n_{i/e}$ is the ion/electron number density, and $\lambda_{i/e}$ is the mean free path of the ions/electrons. These are colloquially known as the ``\textit{Spitzer}'' viscosity and conductivity.

If thermal conductivity were this strong in the ICM, the sharp temperature gradients observed in cold fronts should smear out on very fast timescales. Another consideration is the velocity shear predicted to exist across cold fronts. Such situations are susceptible to the \textit{Kelvin–Helmholtz instability (KHI)}, which occurs at a sharp interface between two fluids moving at different velocities (such as in cold fronts). Small velocity perturbations are amplified into large ripples and waves, which would be seen in X-ray images of clusters. If viscosity were at the Spitzer level in the ICM, it should prevent the growth of this instability. KHI would also be suppressed by a strong magnetic field oriented parallel to the front surface. As we will see in Section \ref{sec:plasma_physics}, the presence of the magnetic field in the ICM changes this overall physical picture substantially, and observations indicate that viscosity and conductivity may not be as strong as predicted by Equations \ref{eqn:visc}-\ref{eqn:cond}.

From the observational perspective, cold fronts are usually grouped into two types, ``merger-remnant'' and ``sloshing'' cold fronts, both of which are described in detail below. Another possible mechanism for forming cold fronts is the collision of two shocks or gas streams,  \cite{Birnboim2010,Zinger2018,Zhang2021}, which will not be discussed in detail here.

\subsubsection{``Merger-Remnant'' Cold Fronts}\label{sec:merger_cf}

When a cold and dense substructure plunges through a hotter and more diffuse medium, a \textit{merger-remnant} cold front can form at the leading edge of the substructure as a result of \textit{ram pressure} confinement of the infalling substructure. The concept of ram pressure is simple and intuitive: the pressure experienced by an object moving relative to a medium. Its strength $P_{\rm ram}$ can be calculated as
\begin{equation}
    P_{\rm ram}=\rho v^2,
\end{equation}
where $v$ is the velocity of a substructure through the ICM of the main cluster and $\rho$ is the density of the ICM. 

In the context of cluster mergers, the substructure can be a subcluster, a galaxy group, or an early-type galaxy, with a cool core or a cooler medium, traveling through the hotter ICM of the main cluster.  

\begin{figure*}
\centering
	\includegraphics[width=0.5\textwidth]{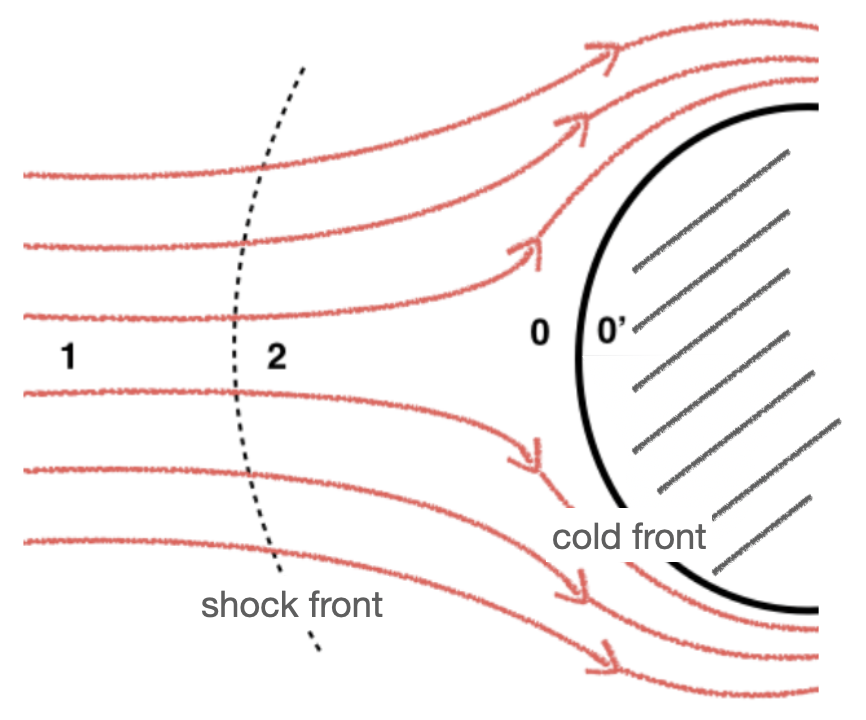}	
    \caption{Scheme of flow past a spheroid. Zones 0, 1, and 2 are the stagnation
point, the undisturbed free stream, and a (possible) post shock region,
respectively. Zone 0’ is within the body. Taken from \cite{Su2017}.}
    \label{fig:flow}
\end{figure*}

The flow structure for such a scenario is displayed in Figure~\ref{fig:flow}. Zone 1 represents the undisturbed {\it free stream}. The region just outside the infalling sphere, along the central axis (Zone 0) is the {\it stagnation point}. The edge that separates Zones 0 and 0' is the cold front. Zone 0' is denser (X-ray brighter) and cooler while Zone 0 is hotter and more diffuse. In observations, it is not practical to directly measure the pressure of Zone 0 but it is much easier to measure that of Zone 0'.
A pressure equilibrium of $P_{0}=P_{0'}$ is generally assumed across a cold front.
The infalling velocity can be calculated using the pressure ratio of the free stream and the stagnation point \citep{Vikhlinin2001, Landau1959}:

\begin{equation}
\frac{P_0}{P_1}=\left(1+\frac{\Gamma-1}{2}{\cal M}\right)^{\Gamma/(\Gamma-1)}, ~{\cal M}\leq 1,
\end{equation}
\begin{equation}
\frac{P_0}{P_1}=\left(\frac{\Gamma+1}{2}\right)^{(\Gamma+1)/(\Gamma-1)}{\cal M}^2\left(\Gamma-\frac{\Gamma-1}{2{\cal M}^2}\right)^{-1/(\Gamma-1)}, ~{\cal M}> 1,
\end{equation}
where $\Gamma=5/3$ is the adiabatic index (ratio of specific heats) of the monatomic gas, and {\cal M} is the Mach number (defined in Equation~\ref{eq:mach}). 
Here, we describe two merger-remnant cold fronts that are very nearby and have extensive X-ray coverage. 

\begin{figure*}
\centering
	\includegraphics[width=0.3975\textwidth]{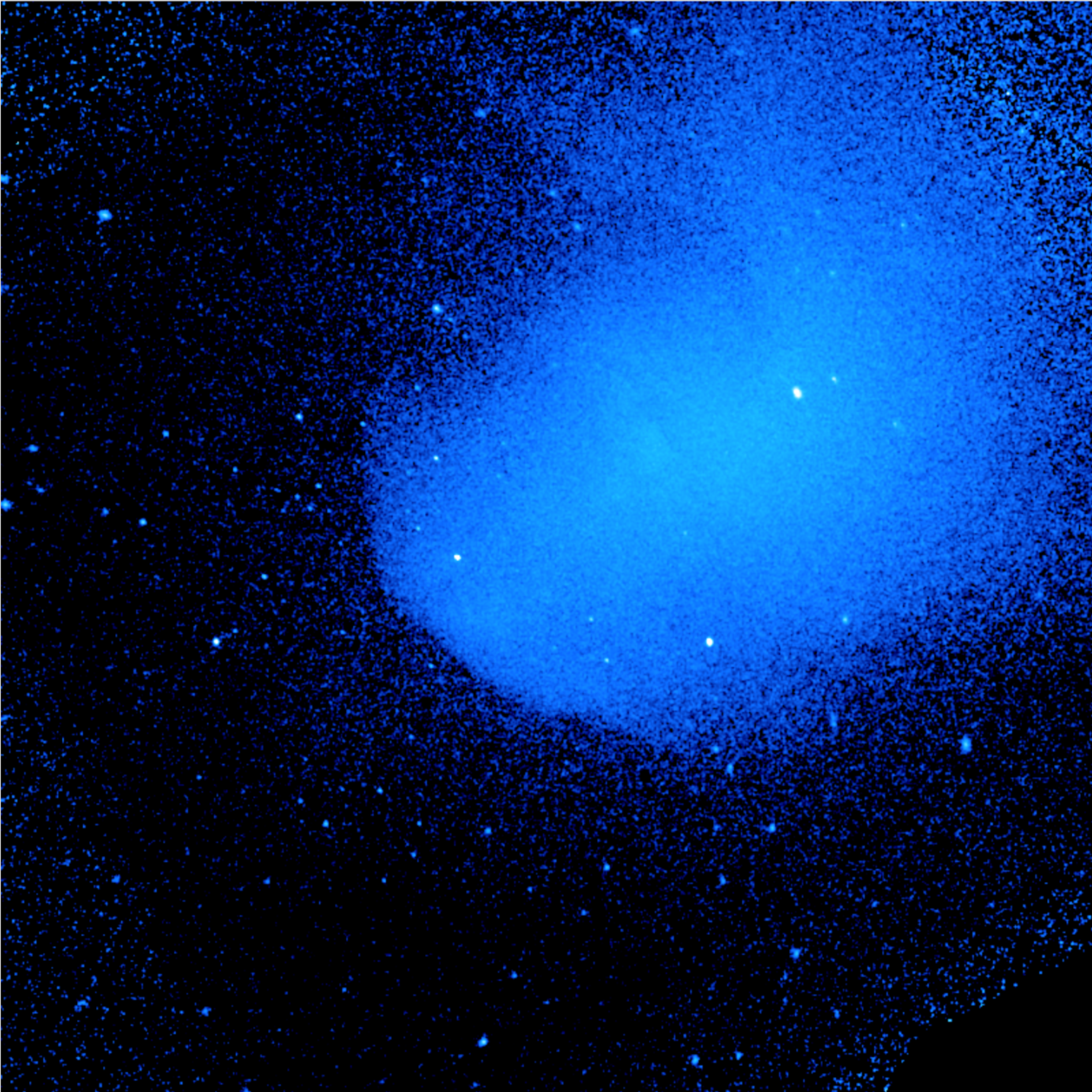}
	\includegraphics[width=0.4\textwidth]{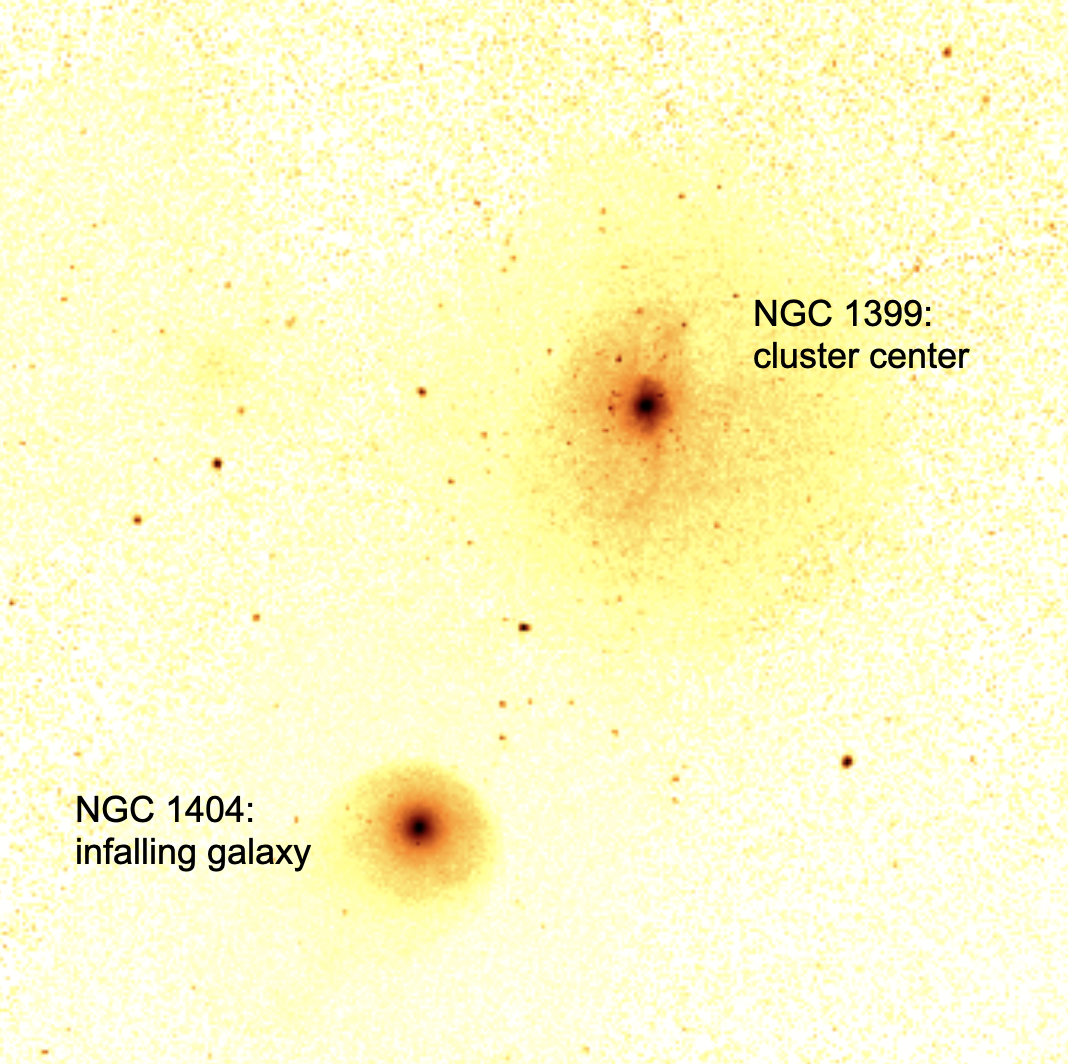}
	\caption{Examples of remnant-core cold fronts. {\it Left:} \textit{Chandra} X-ray image of Abell~3667 \citep{Ichinohe2017}. {\it Right:} \textit{Chandra} X-ray image of the Fornax cluster and the infalling elliptical galaxy NGC~1404 \citep{Su2017}.}
	  \label{fig:rccf}
\end{figure*}

{\bf Abell~3667} is a nearby merging cluster at $z=0.055$. The infalling subcluster has already passed through the center of the main cluster and forms a bright cold front to the southeast as shown in the left panel of Figure~\ref{fig:rccf}. The temperature is $\sim4$\,keV within the cold front, and jumps to 8\,keV just outside. The motion of Abell~3667 has a Mach number of ${\cal M}\lesssim1$. The width of the cold front has been determined to be $\sim$5\,kpc \cite{Vikhlinin2001}, several times smaller than the Coulomb mean free path, implying a suppression of particle diffusion and thermal conduction across this interface, likely due to magnetic fields.
Deep \textit{Chandra} observations reveal unevenness in the
surface brightness at the cold front as well as azimuthal variations, likely due to the presence of KHI developing
along the cold front \cite{Ichinohe2017}, an analysis of which suggests that the viscosity in the Abell~3667 ICM is suppressed to $\sim$5\% of the Spitzer value.

{\bf NGC~1404} is an early-type galaxy falling through the ICM of the Fornax cluster at a distance of $\lesssim20$\,Mpc. A prominent cold front is formed between the 0.6\,keV interstellar medium (ISM) and the 1.5\,keV Fornax ICM \cite{Su2017,Su2017b} as shown in the right panel of Figure~\ref{fig:rccf}. 
No broadening in the surface brightness profile larger than the instrumental
spatial resolution is found across the cold front, which is smaller than the mean free path, implying suppressed electron diffusion. Sub-kpc scale eddies have been detected on the cold front surface \cite{Su2017b} that are cooler than the surrounding ICM, likely generated by KHI, which puts an upper limit on the isotropic viscosity\footnote{Here and throughout this chapter, ``isotropic'' refers to processes with no dependence on spatial direction, whereas ``anisotropic'' typically means dependence of the process with respect to the local magnetic field line direction, see Section \ref{sec:plasma_physics} for more details.} of the ICM of $\sim$5\% of the Spitzer value. A deep \textit{Chandra} observation allows the pressure gradient along its upstream edge to be measured, which has been used to infer the 3D motion of the NGC~1404 through Bernoulli's principle \cite{Su2017}. 
Its probable merger geometry and history have been reproduced in simulations \cite{Sheardown2018}. 

If the infall of the substructure is supersonic, one may expect a bow shock developed in front of the cold front as seen in the examples mentioned in Section \ref{sec:shock_fronts} of 1E 0657-56, Abell~2146, and Abell~520. For an impenetrable obstacle, the standoff distance between the leading edge and the bow shock can be approximated as 
\begin{equation}
D=0.8R\frac{(\Gamma-1){\cal M}^2+2}{(\Gamma+1)({\cal M}^2-1)},
\label{eq:standoff}
\end{equation}    
where $R$ is the radius of a nearly spherical body \cite{Farris1994}. In cluster mergers, the infalling substructure, despite being denser, is not impenetrable, and is generally not spherical. Qualitatively speaking, from Equation~\ref{eq:standoff}, the stronger the shock, the shorter the standoff distance. If the motion is not supersonic ($\mathcal{M}$ approaching 1), the bow shock would be infinitely far away. Therefore, its presence is not expected. 

\subsubsection{``Sloshing'' Cold Fronts}\label{sec:sloshing_cf}

Among the first discoveries of cold fronts by \textit{Chandra}, some appeared in
clusters with bright, cool cores with an apparent absence of merging activity. These manifest as fronts often arranged in a spiral pattern surrounding the cluster core. 
    
A study of the cluster A1795 noted the existence of a cold front and
postulated that the cold front was formed by the \textit{``sloshing''} of the low-entropy gas of the cool
core in the DM-dominated potential well \cite{Markevitch2001}. This hypothesis was first tested with binary merger simulations of clusters extracted from a cosmological simulation with high
mass ratios and nonzero impact parameters \cite{Tittley2005}. The interactions between the clusters set off oscillations of the DM and gas
cores which produced cold fronts. A more detailed study of this process in a large suite of binary cluster merger simulations with varying mass ratios, impact parameters, and cluster gas properties showed that sloshing cold fronts are formed under quite generic conditions provided that the core gas of the larger cluster is not completely disrupted by the subcluster \cite{Ascasibar2006}. 

The basic process of producing sloshing cold fronts in a cool-core cluster (see \cite{Ascasibar2006} for details) is as follows: a subcluster with a mass at least $\sim$a few times smaller
than the cool-core cluster encounters the latter with a high impact parameter, typically on the
order of 0.5-1~Mpc. At first core passage, the subcluster gravitationally accelerates the stellar,
gas, and DM components of the main cluster core. At the same time, the gas core of the main cluster feels the  influence of the ram pressure of the surrounding medium it has been
accelerated into, which the DM core does not feel, eventually causing a spatial separation between the two. As
the subcluster leaves the core region, its influence on the main cluster's core wanes, and the cold,
low-entropy gas falls back into the potential minimum of the cluster, overshooting it, and setting
off a series of sloshing motions which produce cold fronts. Since the subcluster typically approaches with a non-zero impact parameter, the angular momentum of the encounter is transferred to the core gas via interaction with the gas associated with the subcluster, ensuring the sloshing motions produce spiral-shaped cold fronts. 

\begin{figure*}
\centering
\includegraphics[width=0.95\textwidth]{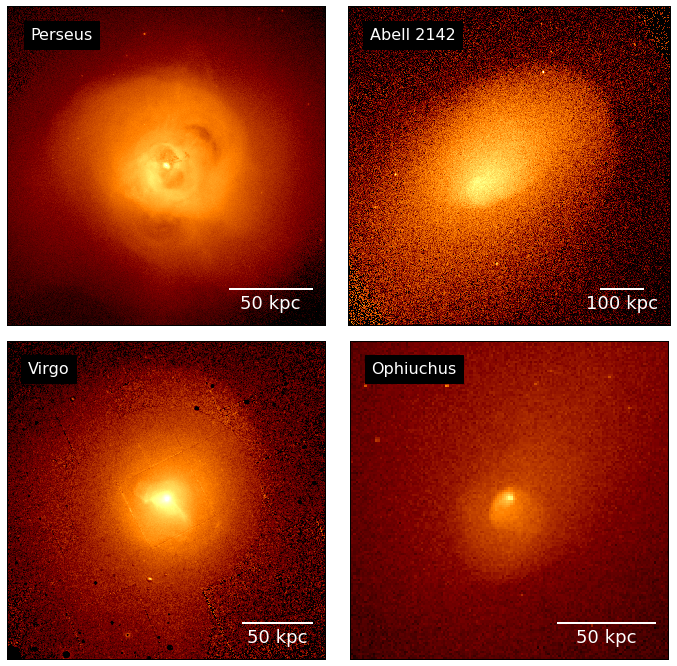}	
\caption{X-ray surface brightness images of four clusters with prominent sloshing cold fronts. Clockwise from upper-left: Perseus (image adapted from \cite{Zhuravleva2014}), Abell 2142 (image adapted from \cite{Wang2018}), Ophiuchus (image adapted from \cite{Giacintucci2020}, and Virgo (image adapted from \cite{Simionescu2010}).}
\label{fig:sloshing_clusters}
\end{figure*}

Some prominent examples of sloshing cold fronts include (all of which are shown in Figure \ref{fig:sloshing_clusters}): 

{\bf Perseus} is a well-studied bright and nearby cluster of galaxies, with extensive observations by \textit{ROSAT}, \textit{Chandra}, \textit{XMM-Newton}, \textit{Suzaku}, and \textit{Hitomi}. Though Perseus is mostly known for the X-ray cavities which are signatures of AGN feedback, it also has a stunning set of sloshing cold fronts \cite{Churazov2003,Fabian2011,Simionescu2012,Walker2017}. Evidence of Kelvin-Helmholtz instabilities have been found along one front \cite{Walker2017}, which can be used to constrain the magnetic field strength in the core by comparing to simulations (see Section \ref{sec:bfields}). The \textit{Hitomi} observations of a bulk velocity gradient across the Perseus core \cite{Hitomi2016,Hitomi2018} likely originates from the same sloshing motions that produce the cold fronts (see also Section \ref{sec:velocity}). These motions may also be responsible for the positions and shapes of the outermost AGN cavities \cite{Fabian2022}.

Perseus is also a good example of the existence of sloshing cold fronts at large radii. A cold front at a radius of $\sim$730~kpc (half the virial radius) was observed by \textit{XMM-Newton} \cite{Simionescu2012}. A deep \textit{Chandra} observation of the same cold front showed that this cold front has split into two fronts, which can serve as a constraint on the magnetic field strength in this region (see \cite{Walker2018} and Section \ref{sec:bfields}).

{\bf Abell 2142} was one of the first clusters observed by \textit{Chandra} to possess cold fronts, which was initially thought to be of the ``merger-remnant'' type discussed in Section \ref{sec:merger_cf} above \cite{Markevitch2000}, but are now believed to be sloshing cold fronts. From these observations, the first constraint on thermal conduction across a cold front was determined to be at least $\sim$250-2500 times below the Spitzer value (see \cite{Ettori2000} and Section \ref{sec:conduction}). Like Perseus, Abell~2142 has a cold front at large radius, $\sim$1~Mpc \cite{Rossetti2013}. A long-exposure \textit{Chandra} study of the cluster showed evidence for Kelvin-Helmholtz instabilities at one of the inner cold fronts as well as potential evidence for an \textit{X-ray channel}, which is a long and thin dip in X-ray surface brightness that is theorized to be a region of enhanced magnetic pressure which displaces a small amount of the thermal gas (see \cite{Wang2018} and Section \ref{sec:bfields}).

{\bf Virgo} is another nearby cluster, hosting the active galaxy M87 at its center. Sloshing cold fronts were discovered in Virgo with \textit{XMM-Newton} and \textit{Suzaku} \cite{Simionescu2007,Simionescu2010}. These cold fronts were reproduced in tailored simulations \cite{Roediger2011}, followed on by simulations including viscosity \cite{Roediger2013,ZuHone2015a} (see Section \ref{sec:viscosity}). A deep \textit{Chandra} observation of the NW cold front used the sharpness of the discontinuity to constrain viscosity and thermal conduction, as well as discover puzzling linear features which may result from amplified magnetic fields (see \cite{Werner2016a}, Figure~\ref{fig:virgo_cold_front}, and Section \ref{sec:bfields}).  

The {\bf Ophiuchus} cluster has a truncated cool core \cite{Werner2016b} and a number of prominent sloshing cold fronts \cite{Million2010,Werner2016b}. The region within the cold fronts has a radio mini-halo and a possible detection of X-ray emission associated with inverse-Compton scattering of cosmic microwave background photons from cosmic ray electrons (see Section \ref{sec:cosmic_rays}). Ophiuchus also has a puzzling concave surface brightness discontinuity to the SE of the core, possibly arising from a large 200~kpc-wide AGN-blown cavity \cite{Werner2016b,Giacintucci2020}, which would be the most powerful AGN explosion known, and may be influenced by the sloshing motions or may have even contributed to them. 

\subsection{Shock Fronts}\label{sec:shock_fronts}

\textit{Shock fronts} are propagating disturbances in a fluid with abrupt jumps in density, temperature, and pressure traveling at a speed faster than the local speed of sound. In the ICM, shock fronts can have multiple origins and varieties. Shocks can be produced at cluster centers via the outbursts of supermassive black holes, which provide a viable way to balance the radiative cooling at the centers of cool core clusters (see Chapter ``AGN Feedback in Groups and Clusters'' for details). Shocks can be created beyond a cluster's virial radius by substructures accreted from cosmic filaments, the so-called {\it accretion shocks}. They are expected to be very strong shocks, but the low density of cluster outskirts makes them essentially undetectable by current X-ray telescopes (more details on these shocks are presented in the Chapter ``Cluster Outskirts and their Connection to the Cosmic Web''). Here we focus on {\it merger shocks}, which are created via the collision of (at least) two subclusters.

When a subcluster falls through the gravitational potential of a main cluster, the gaseous component of the subcluster experiences ram pressure and deceleration. Supersonic gas motion can lead to a {\it bow shock}. 
In X-ray observations, shocks are characterized by a discontinuity in the gas pressure with a bright surface brightness edge associated with an elevated temperature.
The strength of a shock can be described by the \textit{Mach number}, which is the ratio 
of the flow velocity to the sound speed of the medium 
\begin{equation}
    {\cal M}\equiv v/c_s
    \label{eq:mach} 
\end{equation}
and 
the sound speed, $c_s$ can be calculated through 
\begin{equation}
    c_s=\sqrt{\frac{\Gamma k_BT}{\mu{m_p}}},
\end{equation}
where $T$ stands for the thermal temperature of the medium, $\Gamma = 5/3$ is the specific heat  
for a monatomic gas, $k_B$ is the Boltzmann constant, $m_p$ is the proton mass, and the average molecular weight $\mu \approx 0.6$. 
Gas properties across a (plane-parallel) shock are expected to obey the Rankine-Hugoniot jump conditions, describing the conservation of mass, momentum, and energy, respectively:

\begin{equation}
    \rho_1v_1=\rho_2v_2
  \end{equation}
  \begin{equation}
    P_1+\rho_1v_1^2=P_2+\rho_2v_2^2
\end{equation}
  \begin{equation}
 \frac{1}{2}v_1^2+\frac{\Gamma}{\Gamma-1}\frac{P_1}{\rho_1}=\frac{1}{2}v_2^2+\frac{\Gamma}{\Gamma-1}\frac{P_2}{\rho_2}
\end{equation}
where $P$, $\rho$, and $v$ are gas pressure, density, and flow velocity, respectively, with the subscripts 1 and 2 denoting the upstream (before) and downstream (after) regions of the shock and $\rho_2>\rho_1$, $v_2<v_1$, $T_2>T_1$, $P_2>P_1$. One can derive the Mach number through the density or temperature ratios \citep{Landau1959}:
\begin{equation}
{\cal M}=\left[\frac{2{\rho_2/\rho_1}}{\Gamma+1-{\rho_2/\rho_1}(\Gamma-1)}\right]^{1/2}
\end{equation}

\begin{equation}
{\cal M}=\left[\frac{(\Gamma+1)^2({T_2/T_1}-1)}{2\Gamma(\Gamma-1)}\right]^{1/2}
\end{equation}
Additionally, the Mach number can be probed through the angle of the shock front relative to the symmetry
axis (\textit{Mach cone}): 
\begin{equation}
{\cal M}=1/\sin\psi
\end{equation}

Despite that mergers are ubiquitous among galaxy clusters, confirmed merger shocks are rare. They are most easily detected near cluster centers where it is sufficiently X-ray bright\footnote{The ICM is optically thin with a typical temperature of a few keV and an electron density of $\approx10^{-3}$\,cm$^{-3}$. The electron density, however, declines quickly with radius, from $\approx10^{-1}-10^{-2}$\,cm$^{-3}$ near cluster centers to $\approx10^{-4}$\,cm$^{-3}$ at the outskirts. The emissivity of thermal bremsstrahlung is proportional to the square of the gas density. The central region of a cluster can be $\approx10^4 \times$ X-ray brighter than its outskirts.}. 
Merger shocks propagate outward to cluster outskirts at supersonic speed. The time for a shock front to be present near the cluster center is relatively short. 
These measurements are also sensitive to the merger geometry and projection effects.
To date, merger shocks have been detected in more than 20 clusters (e.g., RX J0751.3+5012 \cite{Russell2014}, Abell~754 \cite{Macario2011}, Abell~2256 \cite{Ge2020}, Abell~115 \cite{Botteon2016}, Abell~2219 \cite{Canning2017}, 
3C 438 \cite{Emery2017},
Abell~665 \cite{Dasadia2016}
). 
Here we showcase three clusters in which shocks have been well-studied in X-rays. Some of the concepts will be detailed in subsequent sections. 

\begin{figure*}
\centering
	\includegraphics[width=0.8\textwidth]{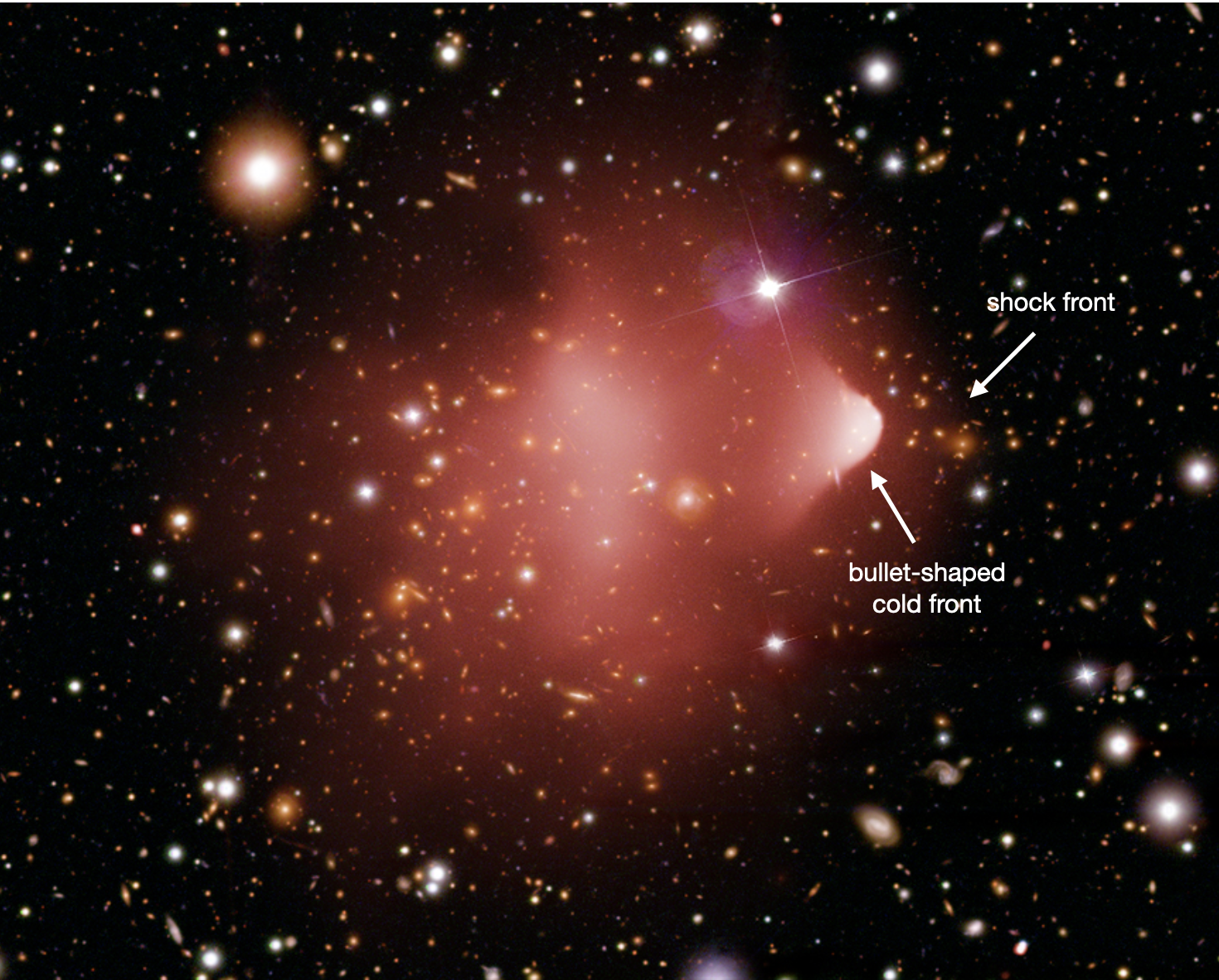}	
    \caption{X-ray and optical composite image of 1E 0657-56. X-ray: NASA/CXC/CfA/M.Markevitch et al.; Optical: NASA/STScI; Magellan/U.Arizona/D.Clowe et al.}
    \label{fig:bullet}
\end{figure*}

The first clear example of a cluster bow shock is found in {\bf 1E 0657-56}, dubbed the ``Bullet Cluster'', at $z=0.296$.
It was first discovered as an extended X-ray source by the Einstein Observatory Image Proportional Counter (IPC) \cite{Tucker1995}. \textit{ROSAT} observations indicated that it is a merging cluster \cite{Tucker1998}. 
As shown in Figure~\ref{fig:bullet}, the \textit{Chandra} observation of 1E 0657-56 highlights a ``bullet" just passing the cluster core and moving towards the west \cite{Markevitch2002}. The ``bullet'' subcluster is preceded by a sharp X-ray brightness edge. Temperature analysis indicates that the gas outside the edge is cooler than that inside, such that the pressure is higher within the edge, providing strong evidence for a shock front. 
The density jump is estimated to be $3.2\pm0.8$ at the shock front, corresponding to a Mach number of $3.4$. The measured temperature jumps from 8\,keV to 18\,keV, corresponding to a 
Mach number of $2.1$\footnote{ 
The discrepancies in the derived Mach number between the density and temperature measurements are common in X-ray observations (see Figure~\ref{fig:wittor}-left), which may be caused by their different dependencies on the viewing angle.}.
About 100 kpc behind this shock front, there is a cold front with no pressure jump, capturing the motion of the remnant cool core of the subcluster (see Section~\ref{sec:merger_cf}).
The Australia Telescope Compact Array observation of 1E 0657-56 at 1.4 GHz shows a Mpc-scale radio relic (see Section \ref{sec:cosmic_rays}) to east of the main cluster, the opposite direction of the X-ray bow shock \cite{Shimwell2015}. An X-ray brightness edge is also observed at the same location, suggesting that this is likely to be a shock front, despite the large uncertainty in the temperature measurement. 

1E 0657-56 is perhaps most famous for providing what has been called ``smoking gun'' evidence of weakly-interacting DM.
During the encounter between two subclusters, their DM halos and member galaxies interact essentially only gravitationally. In contrast, their gaseous components (the ICM) experience ram pressure from the encounter, which slows them down compared to the stars and DM. A joint weak-lensing and X-ray analysis of 1E 0657-56 reveals the spatial separation between the DM and the ICM projected on the sky and puts constraints on the DM self-interaction cross-section \cite{Markevitch2004,Randall2008}.

\begin{figure*}
\centering
	\includegraphics[width=0.8\textwidth]{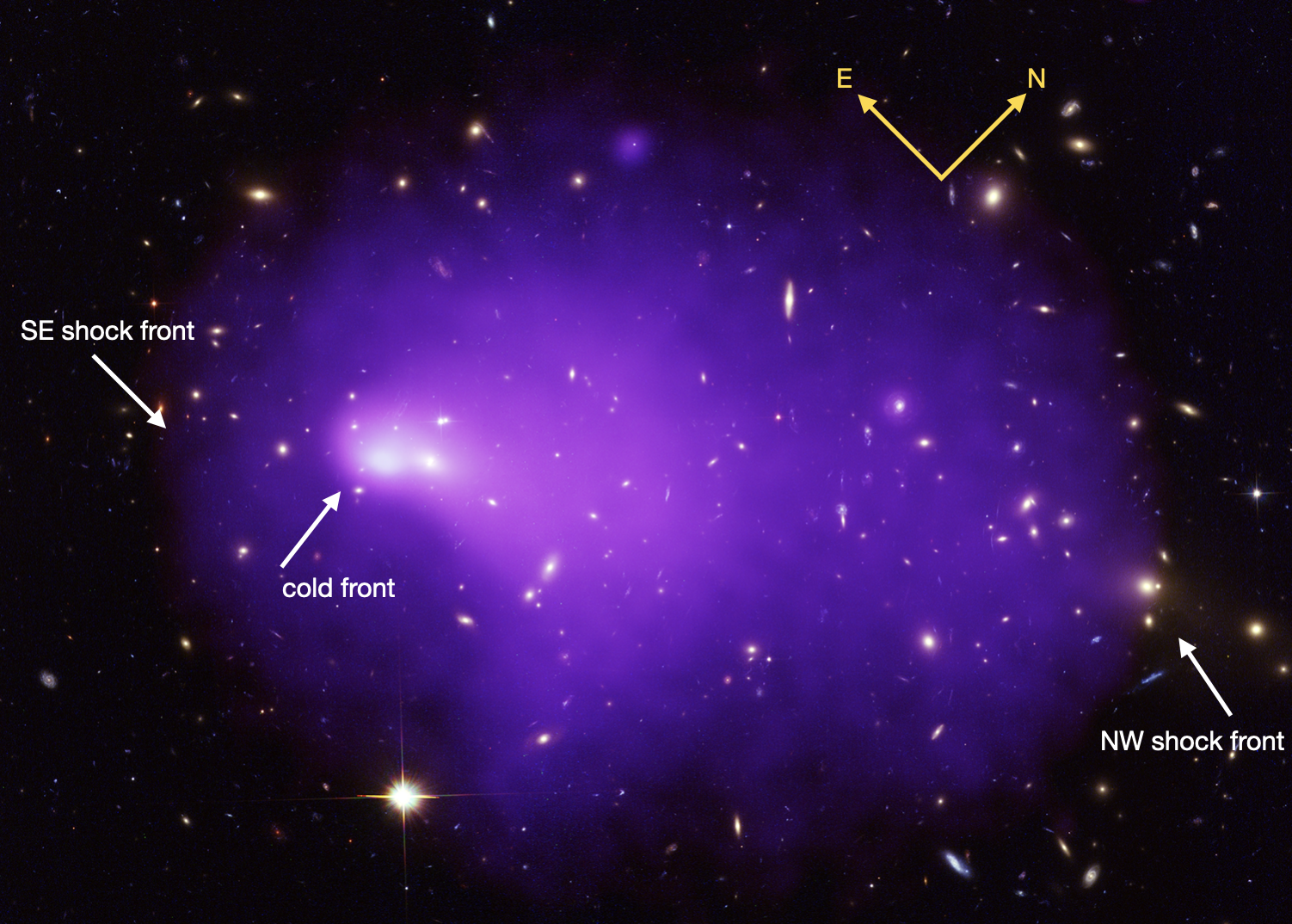}	
    \caption{X-ray (\textit{Chandra}) and optical (HST) composite image of Abell~2146. (Credit: X-ray: NASA/CXC/Univ of Waterloo/H. Russell et al.; Optical: NASA/STScI)}
    \label{fig:A2146}
\end{figure*}

Another cluster that shows two shock fronts
is {\bf Abell~2146}, a galaxy cluster at $z=0.234$.  One bow shock to the southeast and one upstream shock to the northwest \cite{Russell2010} are visible in Figure~\ref{fig:A2146}. The presence of the two shock fronts on opposite sides of the system is a strong indication that the merger is nearly in the plane of the sky such that projection effects are minimized.
The Mach number of the SE bow shock is ${\cal M}=2.2\pm0.8$, corresponding to a shock velocity of $v=1400-3200$\,km\,s$^{-1}$. The subcluster may have passed the center of the main cluster 0.1--0.3\,Gyr ago.
A large pressure jump is observed across the shock front and
a prominent cold front (see Section~\ref{sec:merger_cf}) with no pressure change is observed $\sim150$\,kpc behind this bow shock. 
The cold front is formed as
the dense cool core of the subcluster passes through
the center of the main cluster and is confined by ram pressure. The morphology of the remnant cool core of the subcluster implies that the cool core is being destroyed (or will be soon) by the collision. 
The upstream (NW) shock is propagating slower towards the opposite direction with a Mach number of ${\cal M}=1.7\pm0.3$. These features of the cluster are reproduced well by the tailored simulations of \cite{Chadayammuri2022}.
The multiconfiguration 1-2 GHz Jansky Very Large Array (JVLA) observations of Abell~2146 also revealed multiple extended radio components. In particular, the one associated with the upstream shock has been classified as a radio relic (see \cite{Hlavacek-Larrondo2018} and Section \ref{sec:cosmic_rays}).

\begin{figure*}
\centering
	\includegraphics[width=0.8\textwidth]{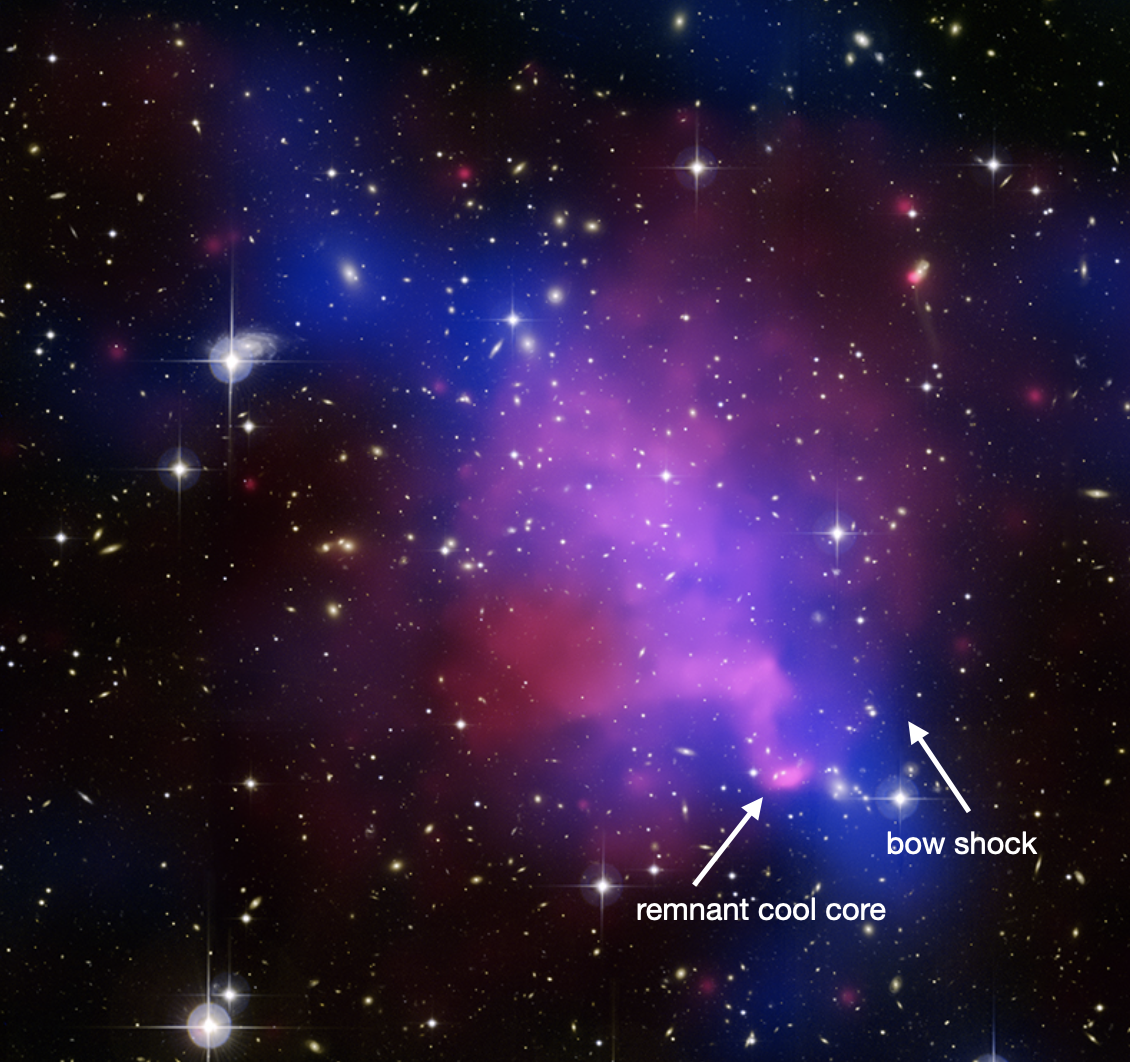}
	 \caption{Composite image of Abell~520. Hot gas distribution is shown in magenta. Mass distribution determined through weak-lensing is shown in blue. Credit: X-ray: NASA/CXC/UVic./A.Mahdavi et al. Optical/Lensing: CFHT/UVic./A.Mahdavi et al.}
    \label{fig:A520}
\end{figure*}

{\bf Abell~520} is a merging cluster at $z=0.203$. Its X-ray morphology is more complex than 1E 0657-56 and Abell~2146, as shown in Figure~\ref{fig:A520}. Thus, the merger scenario is less clear-cut. Still, a major merger can be inferred along the northeast-southwest axis from \textit{Chandra} observations \cite{Markevitch2005,Wang2018}. A bow shock is detected  
to the southwest perpendicular to the merger axis. The density jump gives a Mach number of ${\cal M}=2.4^{+0.4}_{-0.2}$. The temperature is significantly higher at the brighter side, which confirms that this is a shock front.
The bow shock is preceded by a remnant of a cool core that is in the process of being disrupted by ram-pressure stripping.  

\begin{figure*}
\centering
	\includegraphics[width=0.9\textwidth]{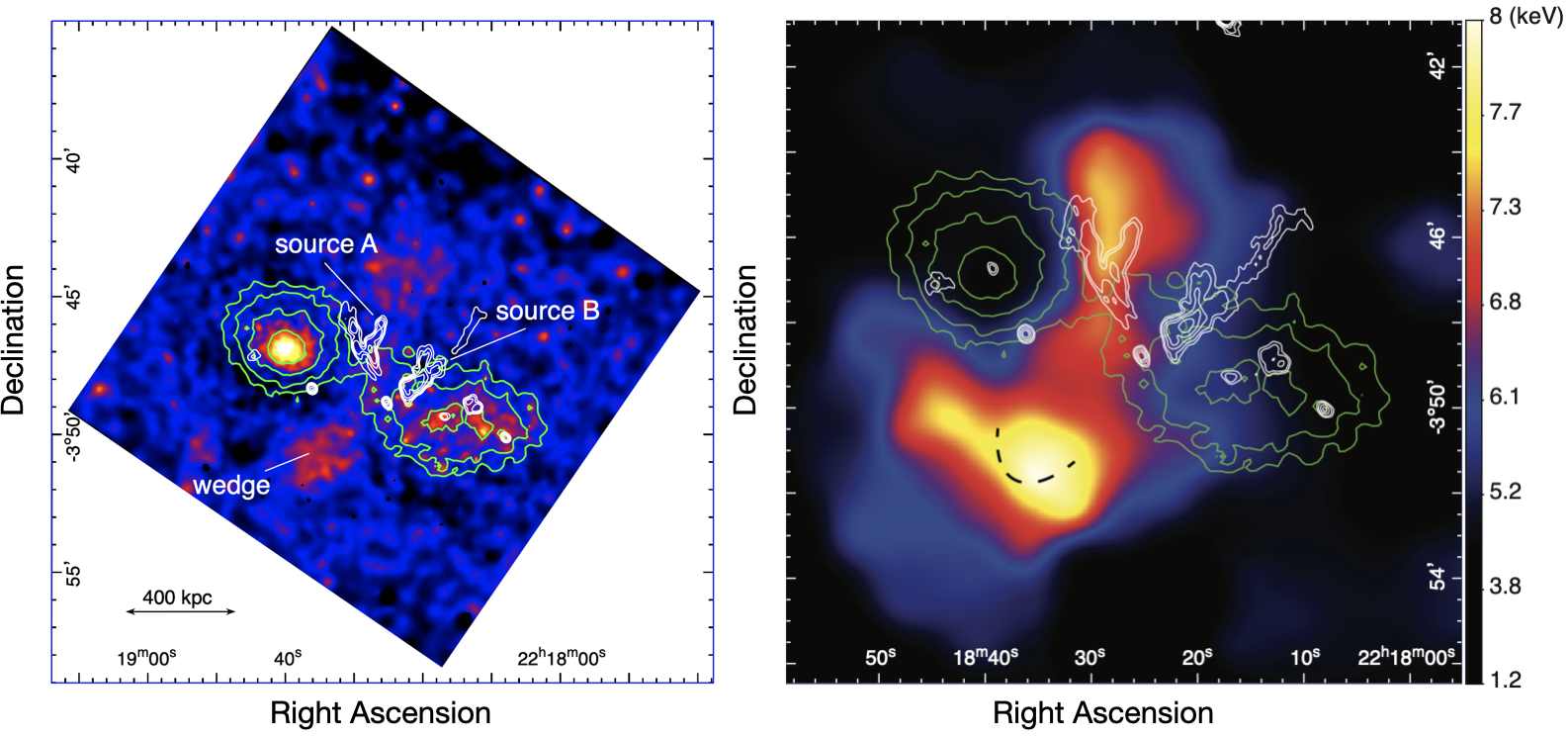}	
    \caption{Residual image and temperature map of the pre-merging cluster 1E 2216.0-0401 and 1E 2215.7-0404 taken from \cite{Gu2019}. The green and white contours trace the X-ray and radio 325 MHz emissions, respectively. 
    {\it Left:} Residual \textit{Chandra} image, created by 
    subtracting a 2D beta model for both clusters,
    highlights a wedge-like feature towards southeast on the merger equator. 
    {\it Right:} ICM temperature map  derived with \textit{XMM-Newton}. The edge of the wedge feature is marked with a black dashed curve. Elevated temperatures are observed at the wedge feature as well as the opposite side of the merge axis.}
    \label{fig:premerger}
\end{figure*}

As noted above, shocks can be best studied near the cluster center (typically at a post-merger phase and shortly after the core passage) where the X-ray surface brightness is sufficiently high such that the shock edge can be detected and the gas temperature on the fainter side of the edge can be constrained. Nevertheless, shocks at large radii of galaxy clusters are expected, and have been detected. For example, theoretical models predict that merger shocks can be created before the core collision via the interaction of the outer atmospheres of two clusters, and propagate outward perpendicular to the merger axis \cite{Ha2018}.
Tentative observational evidence of this effect, often featuring a heated region between two clusters in the early stages of merging, has been found in a number of systems, such as the merger of Abell~399/401 \cite{Akamatsu2017}, Abell~1758 \cite{Botteon2018}, and the accretion of M49 onto the Virgo cluster \citep{Simionescu2017, Su2019}. However, no clear surface brightness edges have been detected in these systems, making it difficult to distinguish between the possibilities of a pre-merger shock and simply adiabatic compression\footnote{A shock involves supersonic motion which results in a discontinuous increase in density, temperature, pressure, and entropy, whereas adiabatic compression increases the density, temperature, and pressure of the gas at constant entropy, and is not associated with any discontinuity.}. {\it Equatorial shock} has been clearly detected for the first time between clusters 1E 2216.0-0401 and 1E 2215.7-0404 {\it before} their collision \cite{Gu2019}. 
As shown in Figure~\ref{fig:premerger}, a wedge-like protruding feature in the X-ray surface brightness resides southeast of the merging equatorial and hundreds of kpc away from the center of each cluster. Elevated temperatures have been detected between the two clusters, particularly associated with the wedge feature. 
Extended radio emission has been detected on the other side (northwest) of the merger axis. The detection of this \textit{pre-merger} shock confirms an important stage in the merging process of galaxy clusters.  

A merger shock propagates away from the main cluster center after the first core passage of a merger. Simulations show that when such a ``runaway merger shock" encounters the accretion shock at cluster outskirts, a new long-lived shock is formed and propagates beyond the virial radii, referred to as a ``merger-accelerated accretion shock'' \cite{Zhang2020}. However, the equilibration time between ions and electrons can be relatively long at cluster outskirts, making the measurement exceptionally challenging (see Section \ref{sec:shock_ei_eq} and the Chapter ``Cluster Outskirts and their Connection to the Cosmic Web'').

\subsection{Ram-Pressure-Stripped Tails}\label{sec:tails}
Cluster galaxies live in a fast lane of galaxy evolution, as seen by the Butcher-Oemler effect---the
excess of optically blue galaxies in clusters at higher redshifts when compared to similar nearby clusters \cite{Butcher1978, Butcher1984}.
Ram pressure stripping is one of the most efficient mechanisms driving the evolution of cluster galaxies \cite{Gunn1972}.
Once the strength of the ram pressure exceeds the gravitational attraction of the infalling galaxy, its ISM can be removed from the stellar body and form a stripped tail trailing behind. For a star-forming galaxy, the cold gas that serves as the fuel for star formation can be removed through ram pressure stripping and eventually ``quench'' the galaxy.
There are ample observations of lopsided, clearly ram pressure stripped tails in HI \cite{Kenney2004,Chung2007,Abramson2011}, H${\alpha}$ \cite{Sun2007,Sun2021,Yagi2007, Yoshida2004, Yoshida2008, Kenney2008}, X-rays \cite{Wang2004, Finoguenov2004,Su2014}, and radio  \cite{Ignesti2022} among cluster galaxies. 

\begin{figure*}
\centering
	\includegraphics[width=0.4\textwidth]{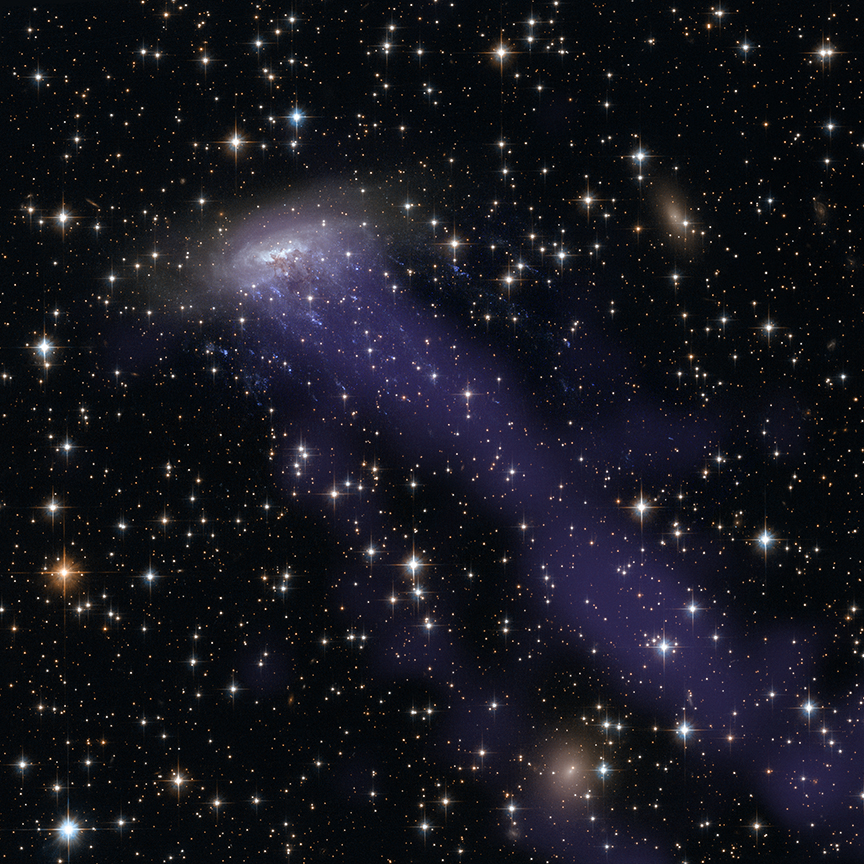}	
		\includegraphics[width=0.45\textwidth]{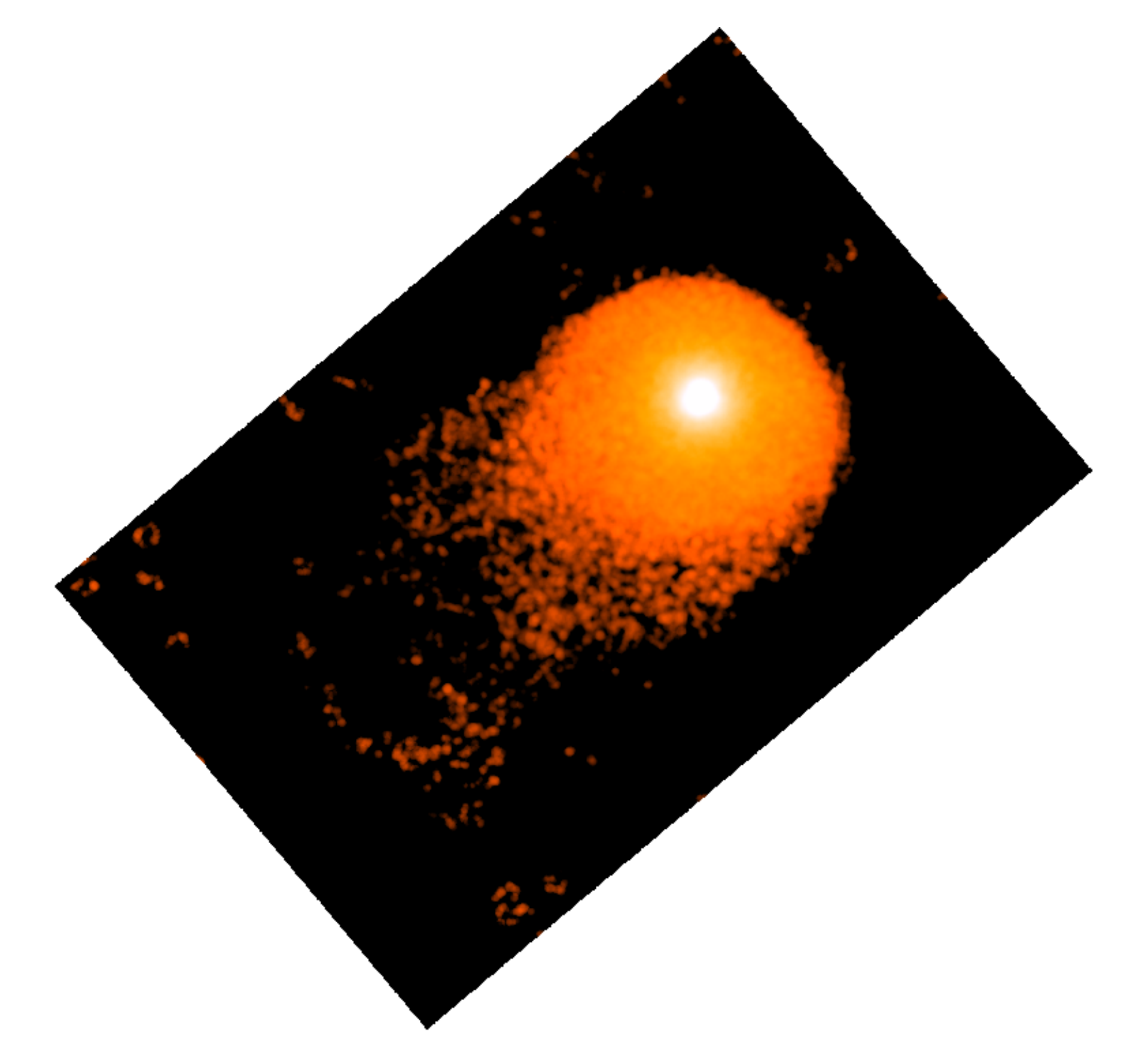}	
    \caption{Examples of ram pressure stripped tails. \textit{Left:} composite image of ESO 137-001 from the Hubble Space Telescope and the Chandra X-ray Observatory. Credit: X-ray: NASA/CXC/UAH/M.Sun et al; Optical: NASA, ESA, and the Hubble Heritage Team (STScI/AURA). \textit{Right:} \textit{Chandra} X-ray image of NGC~1404 adapted from \cite{Su2017}.}
    \label{fig:eso137}
\end{figure*}

A textbook example of ram pressure stripping is ESO 137-001 as shown in Figure~\ref{fig:eso137}-left.
It is a barred spiral galaxy falling through the very massive cluster Abell~3267 \cite{Woudt2008}.
\textit{Chandra} and \textit{XMM-Newton} imaging revealed a long, narrow, and bifurcated X-ray tail extending out to $\sim$70 kpc \cite{Sun2006, Sun2010}.
The bright X-ray tail may form from mixing of cold stripped ISM with the surrounding hot ICM \cite{Sun2010, Tonnesen2011}. 
Optical images of ESO 137-001 reveal a $\sim40$\,kpc long
H${\alpha}$ tail comprised of diffuse gas and a large amount of warm H$_2$ emission \cite{Sun2007}.
The MUSE observations show that the outskirts of ESO 137-001 are already completely devoid of cold gas \cite{Fumagalli2014}.
ESO 137-001 presents an ongoing
transformation from a gas-rich and star forming galaxy to a red and dead
galaxy due to ram pressure
stripping. 

Multi-phase gaseous tails covering from $10$\,K to $10^7$\,K are typically associated with late-type galaxies such as ESO 137-001. The most spectacular examples are ``jellyfish galaxies''\footnote{Infalling galaxies interact with the ICM via ram pressure stripping, triggering starbursts along a tail of gas, which often resemble the morphology of a jellyfish.}. A correlation between the X-ray surface brightness and H$\alpha$ surface brightness of gaseous tails supports the hypothesis of the mixing of the stripped ISM with the hot ICM as the origin of multi-phase gas in ram pressure stripped tails \cite{Sun2021}. Such an environment resembles cluster cool cores (see Chapter ``AGN Feedback in Groups and Clusters"). Stripped tails can also radiate in radio, where the relativistic electrons may stem from supernova or AGN (see Section \ref{sec:cosmic_rays}) and the magnetic field in the tail may be amplified due to \textit{shear amplification} (see Section \ref{sec:bfields}).   

The morphology and thermal structure of the stripped X-ray tails of early-type galaxies have been utilized to probe the microscopic nature of the ICM (see Section \ref{sec:plasma_physics}). 
The stripped tail, being cooler and denser than the surrounding ICM, may not survive long due to KHI or thermal conduction. Simulations show that the stripped gas can mix efficiently with the ambient ICM for an inviscid plasma (see Section \ref{sec:viscosity}). This mixing would be suppressed as viscosity increases, such that an extended and cool tail is expected \cite{Roediger2015a,Roediger2015b}. Note that it is also important to disentangle the effect of merger history in shaping the stripped tail \cite{Su2017,Su2017b}.  
In the case of NGC~1404, the downstream tail appears truncated (even after taking into account projection effects) as shown in Figure~\ref{fig:eso137}-right. The tail temperature 
gradually increases from 
the cooler ISM value to the hotter ICM value, favoring a low-viscosity plasma \cite{Su2017b}.

\begin{figure*}
\centering
	\includegraphics[width=0.3\textwidth]{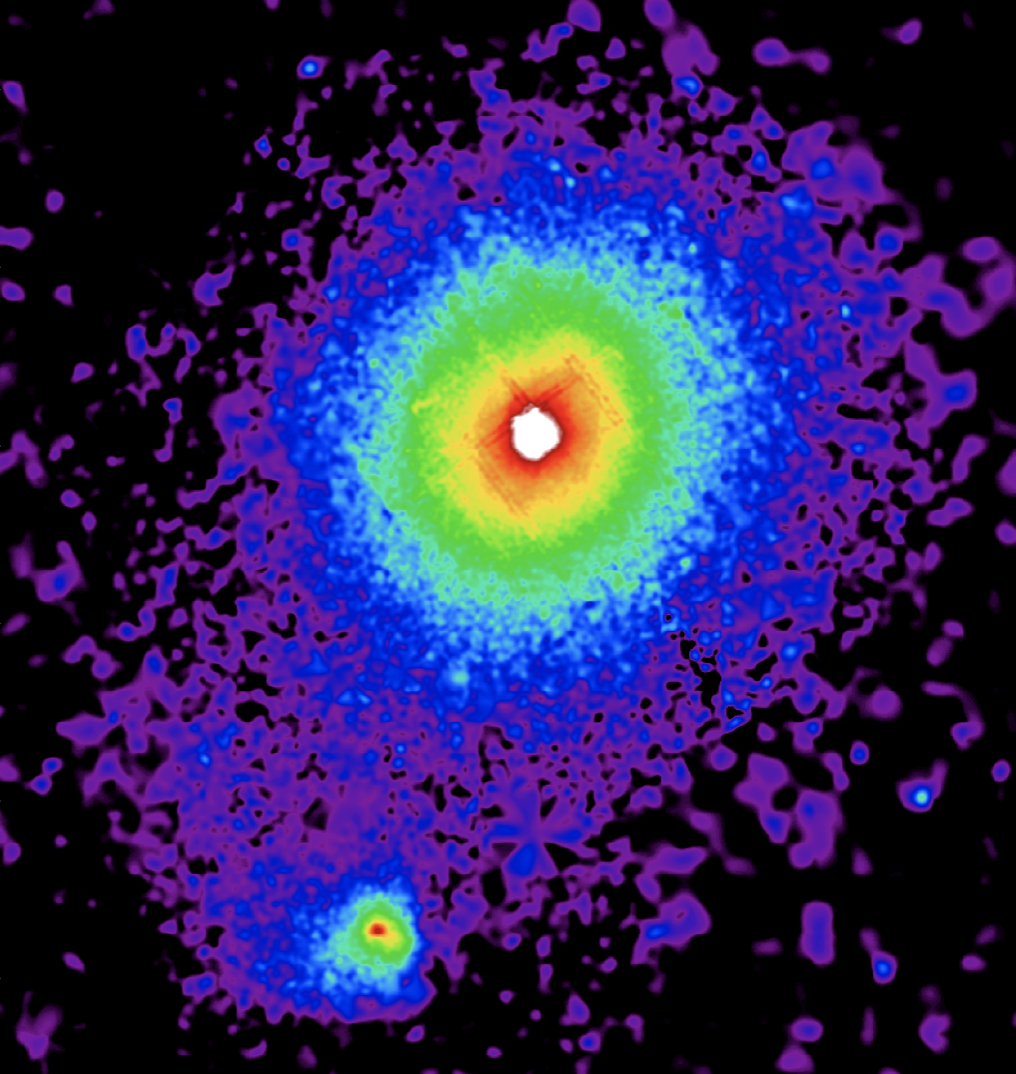}
	\includegraphics[width=0.45\textwidth]{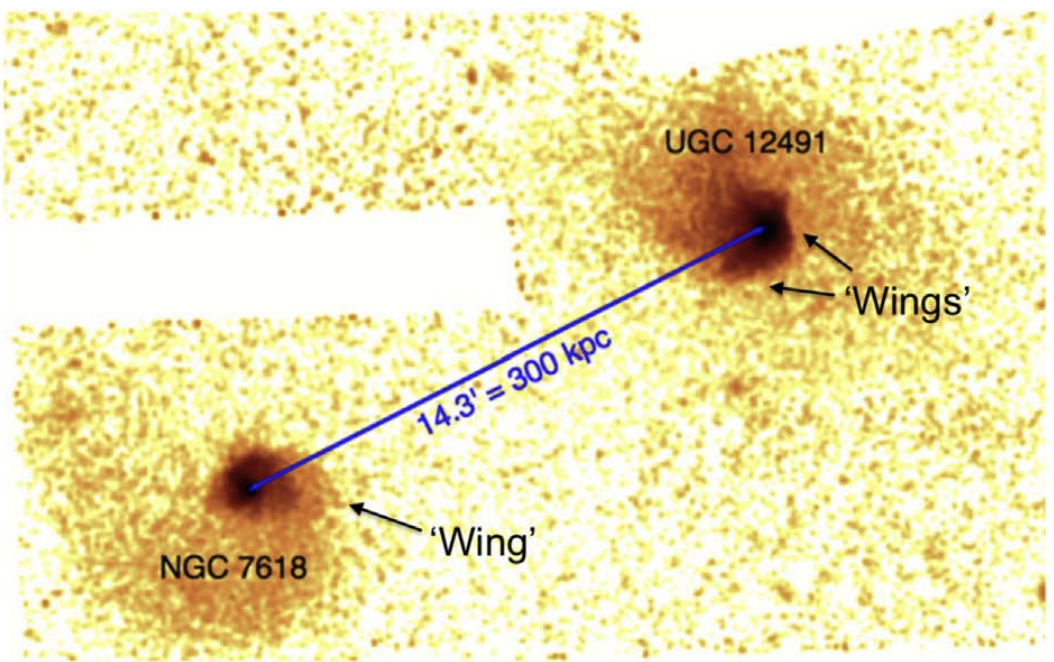}	
    \caption{Examples of slingshot tails. {\it Left:} \textit{XMM-Newton} 
mosaic X-ray image of the Hydra A cluster \citep{DeGrandi2016}. {\it Right:} \textit{Chandra} X-ray image of the NGC 7618 and UGC
12491 galaxy groups \citep{Roediger2012}. }
    \label{fig:slingshot}
\end{figure*}

A relatively new class of X-ray tails are \textit{slingshot tails} \cite{Sheardown2019}. They are formed as a gaseous subhalo (early-type galaxy or galaxy group) passing through a larger galaxy cluster with a large impact parameter approaches the apocenter of its orbit. Unlike typical ram-pressure-stripped tails tracing the recent orbit of the subhalo, slingshot tails can point perpendicular or even opposite to the direction of motion. During the slingshot stage, the tail morphology is influenced by tidal forces rather than just ram pressure. Two typical examples of slingshot tails are shown in Figure~\ref{fig:slingshot}. Both are observed at large distance from their cluster centers.

\section{The Measurement of Merger-Driven Gas Motions}\label{sec:velocity}

Cluster mergers will drive bulk motions and turbulence which can potentially be measured directly by the \textit{shifting and broadening of emission lines} in the cluster's thermal spectrum via the Doppler effect. \textit{Bulk motions} result in Doppler shifts of emission lines, whereas \textit{turbulence} will broaden emission lines. However, line broadening can also be caused by bulk motions if oppositely-directed flows are aligned along the line of sight, which occurs both in cold and shock fronts \cite{ZuHone2016}. The significance of this latter fact for interpreting cluster kinematics will be described below.

The current generation of X-ray telescopes does not typically have the requisite spectral resolution to directly measure motions in the ICM. Upper limits on measurements of bulk motions in clusters were first reported in analysis of \textit{Suzaku} data, in the Centaurus cluster ($\sim$1400~km~s$^{-1}$, \cite{Ota2007}) and Abell 2319 ($\sim$2000~km~s$^{-1}$, \cite{Sugawara2009}). Detections of gas bulk motions were reported from \textit{Suzaku} data in the merging cluster Abell 2256 ($\sim$1500~km~s$^{-1}$, \cite{Tamura2011}) and the Perseus Cluster ($\sim$150-300~km~s$^{-1}$, \cite{Tamura2014}). 

\begin{figure}
\centering
\includegraphics[width=0.95\textwidth]{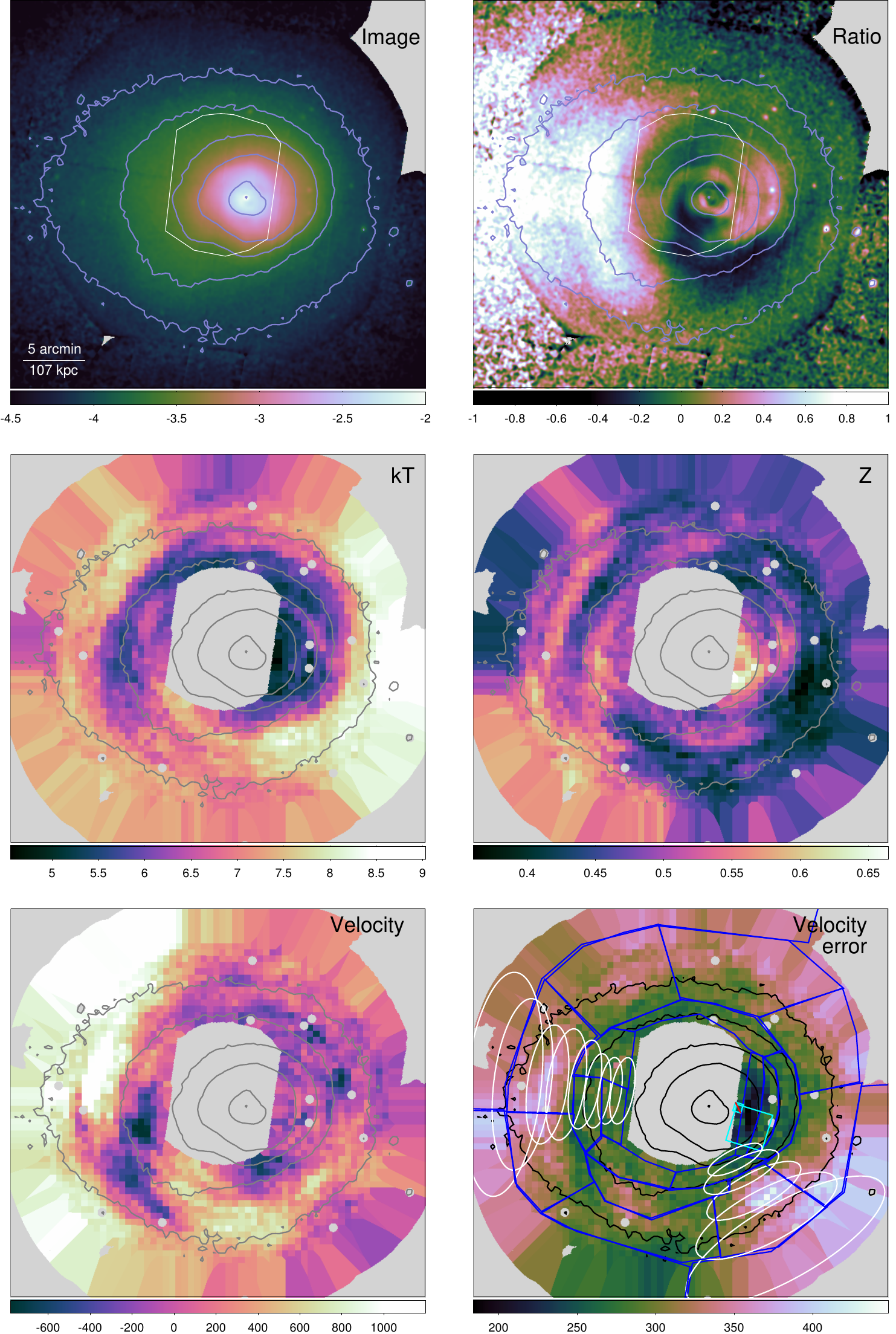}
\caption{Velocities (and other quantities) in the Perseus cluster as measured by \textit{XMM-Newton}, using the fluorescent instrumental background lines. Counter-clockwise from the top-left, the panels are: Surface brightness in the 0.5-2.0~keV band, temperature, line-of-sight velocity shift, error on the velocity shift, metallicity, and fractional difference of surface brightness to radial average. Reproduced from \cite{Sanders2020}.}
\label{fig:xmm_perseus}
\end{figure}

A novel technique to measure ICM velocities uses the fluorescent instrumental background lines present in \textit{XMM-Newton} observations (in particular Cu-K$\alpha$) to calibrate the absolute energy scale of the detector to better than 150~km~s$^{-1}$ at the Fe-K line. Using this technique, \cite{Sanders2020} made velocity maps of Perseus (see Figure \ref{fig:xmm_perseus}) and Coma, both clusters with evidence of ongoing or past merger activity. In Perseus, they found a $\sim$500~km~s$^{-1}$ flow in the location of a sloshing cold front, and in Coma, they found a $\sim$1000~km~s$^{-1}$ range of velocities associated with the different merging components of the system. A similar analysis has been performed on the Virgo cluster \cite{Gatuzz2021}, which showed velocities associated with the cold fronts which are consistent with sloshing, but far in excess of those predicted by numerical models for the Virgo cluster (originally presented in \cite{Roediger2011} and also used in \cite{Roediger2013,ZuHone2015a}), most likely due to the influence of outflows from the central AGN. 

The best spatially and spectrally resolved measurements of ICM velocities require \textit{microcalorimeter} instruments, which are capable of imaging extended X-ray sources such as clusters with $\sim$2-4~eV energy resolution (see Chapter ``TES Detectors'' in ``Section II: Detectors for X-ray Astrophysics''. Assuming that velocities would be best constrained using the He-like Fe-K emission line complex at a rest-frame energy of $E_0 \sim 6.7$~keV (which is at large energy and is relatively isolated from other line complexes), the velocity of the gas may be measured with a precision of $\sim$10s of km~s$^{-1}$. Such a capability would enable the spatially resolved direct measurement of bulk motions and turbulence in many nearby clusters. 

\begin{figure}
\centering
\includegraphics[width=0.95\textwidth]{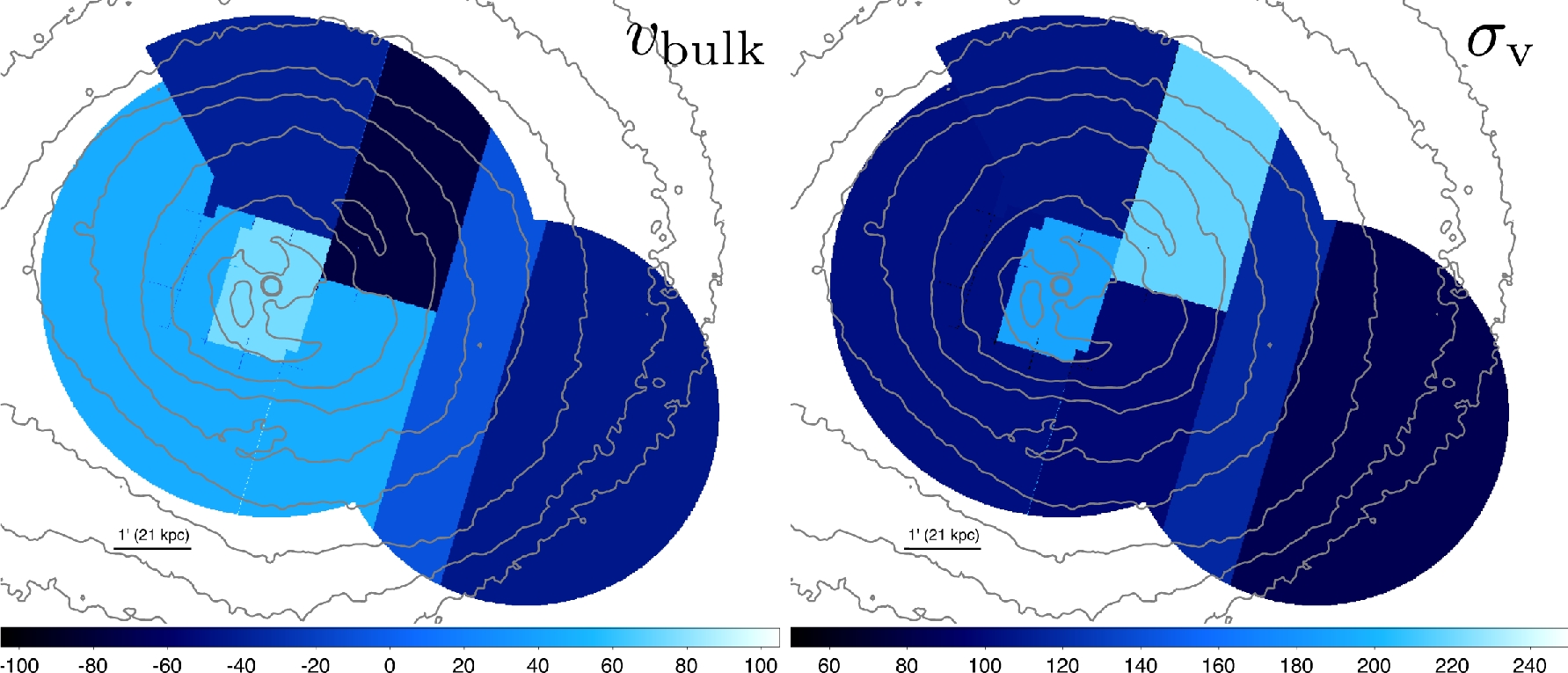}
\caption{Velocities in the Perseus cluster as measured by \textit{Hitomi}. {\it Left}: Bulk velocity map. {\it Right}: Velocity dispersion map. The units in both panels are km~s$^{-1}$. The Chandra X-ray contours are overlaid. Reproduced from \cite{Hitomi2018}.}
\label{fig:hitomi_perseus}
\end{figure}

This capability was briefly available on the \textit{Hitomi} spacecraft, which carried a microcalorimeter instrument. \textit{Hitomi} was launched in February 2016 and was tragically lost in March of the same year. During this brief period, \textit{Hitomi} observed the core of the Perseus cluster, and measured the shifting and broadening of the Fe-K line complex \cite{Hitomi2016,Hitomi2018}. These measurements showed that the core of Perseus turns out to be very ``quiescent'', with a velocity dispersion no larger than $\sim$200~km~s$^{-1}$, which corresponds to a turbulent pressure only $\sim$10\% of the thermal value. The observations also revealed a line-of-sight velocity gradient with a $\sim$100~km~s$^{-1}$ amplitude across the cluster core, consistent with merger-driven gas sloshing (see Figure \ref{fig:hitomi_perseus}). 

Initially, it appears somewhat surprising that such small velocities would be measured in a cluster core with significant AGN activity and evident signatures of gas sloshing. It could be supposed that the reason for this is that the effective viscosity of the Perseus ICM is high. This possibility was tested by \cite{ZuHone2018} with two simulations of gas sloshing in a cluster similar to Perseus; one inviscid, the other highly viscous with isotropic Spitzer viscosity. It was determined that though line shifts and widths on small angular scales are dramatically different from each other in the two simulations, that at the angular resolution of \textit{Hitomi} ($\sim$1') the observed velocity moments are very similar and consistent with each other to within statistical errors and expected differences from cosmic variance. This is because the dominant contribution to Doppler broadening (and spatial differences in Doppler shifting), especially on such large angular scales, comes primarily from velocity differences on scales closer to the injection scale of the motions, rather than the smaller scales primarily affected by viscosity \cite{Inogamov2003}. Because of these limitations imposed by the coarse angular resolution of \textit{Hitomi} (and its successor mission \textit{XRISM}, planned for launch in 2023), spatially resolved measurements of the merger-driven turbulent cascade in nearby clusters await observations by the \textit{Athena} mission and the proposed mission concept \textit{Lynx}, both of which will carry microcalorimeters with much improved angular resolution.      

Gas motions in clusters can also be measured directly via resonant scattering of emission lines, and indirectly via surface brightness fluctuations, both of which are discussed in Chapter ``Plasma Physics of the ICM/IGrM.''

\section{The Impact of ICM Plasma Physics on Merger-Driven Features}\label{sec:plasma_physics}

Most simulations of clusters (whether idealized binary mergers or in cosmological simulations) have assumed that
the ICM can be modeled by ideal hydrodynamics, and many have also included the effects of magnetic
fields. However, other simulations have taken the additional step of exploring dissipative processes
in the ICM; namely its viscosity and the thermal conductivity. As described in Section \ref{sec:cold_fronts}, in the simplest picture Coulomb
collisions set the viscosity and thermal conductivity as a strong function of temperature in what is known as the ``Spitzer'' prescription \cite{Spitzer1962}. However, despite the
fact that the ICM magnetic field is dynamically weak, it nevertheless influences these processes
substantially. The reason for this is that the \textit{Larmor radii} $\rho_L$ of the electrons and ions (the radii of circular motion of charged particles in a local magnetic field) in the ICM are many orders of magnitude smaller than their mean free
paths $\lambda_{\rm mfp}$:
\begin{equation}\label{eqn:larmor}
\rho_L = \frac{mcv_\perp}{|q|B} \sim 10^{-12}~\rm{kpc}
\end{equation}
where $m$ is the mass of the particle, $c$ is the speed of light, $v_\perp$ is the component of the particle velocity perpendicular to the magnetic field, and $B$ is the magnetic field strength. The most probable value for $v_\perp$, assuming a Maxwellian speed distribution, is $v_\perp = \sqrt{2k_BT/m}$. Equation \ref{eqn:larmor} has been calculated for typical ICM conditions and varies only within 2-3 orders of magnitude depending on whether one considers electrons or ions.

This effectively means that viscosity and/or thermal conduction in the ICM should be highly
\textit{anisotropic} with respect to the local magnetic field direction. Depending on the magnetic field
geometry within a given volume in a cluster (whether relatively straight or greatly turbulent), the
effective viscosity and/or thermal conduction can be greatly suppressed relative to the Spitzer value. 

Given this, cold fronts in particular have several characteristics which make them prime candidates for exploring the
microphysics of the ICM plasma by comparing observations to simulations. As mentioned above, their
edges in surface brightness, as observed in high-angular resolution X-ray \textit{Chandra}
observations, are very sharp, with widths smaller than the electron or ion mean free paths---a few
kpc assuming this is set by Coulomb collisions. Such edges should be smeared out by diffusion of
particles at this length scale. Similarly, cold front temperature gradients should be smeared out on
very fast timescales by thermal conduction, assuming again that Coulomb collisions set it at the
Spitzer value. Finally, many cold fronts appear very smooth, but the velocity shear across the
surface of the fronts (which we are unable to observe, but should exist as predicted from
simulations) should make them susceptible to KHI. However, the KHI-causing perturbations will be suppressed in the presence of a stabilizing magnetic field. A significant viscosity will smear out the tangential velocity jump at the front surface and damp growing perturbations. These considerations are clearly affected by the properties of the magnetic field. 

Besides cold fronts, the tails of ram-pressured-stripped galaxies traveling through the ICM provide constraints on viscosity and thermal conduction, and shocks may be used to provide constraints on electron-ion equilibration processes. We will discuss each of these issues in brief in what follows. 

\subsection{Magnetic Fields}\label{sec:bfields}

ICM features such as cold fronts and radio bubbles moving subsonically (${\cal M} \equiv v/c_s < 1$) but super-Alfv\'enically (${\cal M}_A = v/v_A > 1$) though a
magnetized ICM will lead to \textit{magnetic draping}: the formation of a thin, strongly magnetized
boundary layer with a tangential magnetic field \cite{Lyutikov2006}. These two requirements are easily fulfilled in the
ICM, since the magnetic field is dynamically weak with $\beta = p_{\rm th}/p_B \sim 100$. These layers will form under quite generic conditions, regardless
of the magnetic field geometry in the ICM, and that the magnetic pressure in the layer could reach
near-equipartition with the thermal pressure \cite{Lyutikov2006}.

The significance of this layer is that it mitigates the problems mentioned above with a purely
hydrodynamic scenario for cold fronts. The reasons for this are twofold. The first has been already mentioned: given the
small Larmor radii of electrons and ions in the ICM, particle diffusion and thermal conduction
should be very effectively suppressed perpendicular to this draping layer and to the cold front, to
the extent that the layer is effectively draped along the front surface. Secondly, the very strong
magnetic field in the layer and its tangential orientation are able to suppress the growth of
Kelvin-Helmholtz-unstable perturbations along the cold front surface. 

\begin{figure}
\centering
\includegraphics[width=0.95\textwidth]{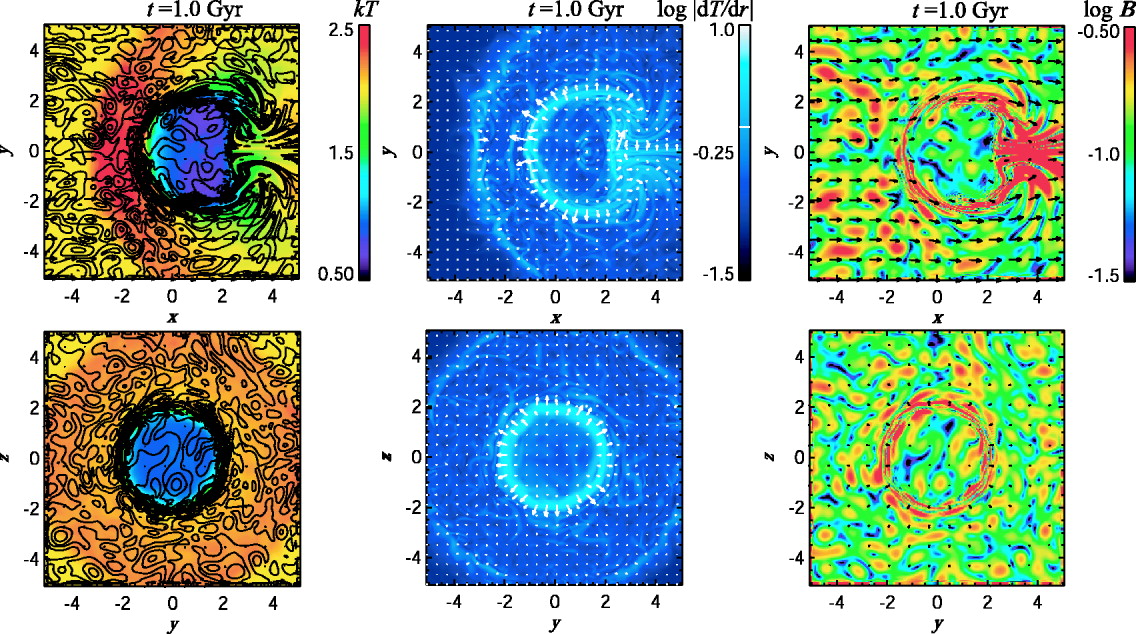}
\caption{A remnant-core cold front propagating through the ICM with a tangled magnetic field in the simulations of \cite{Asai2007}. Panels show slices of temperature (left), temperature gradient (center), and magnetic field strength (right). The two rows show slices along the $z$ (top) and $x$ (bottom) coordinate axes.}
\label{fig:asai_front}
\end{figure}

The formation of this layer was confirmed in simplified simulations of a cold cloud moving in a
hotter, uniform ICM in several works (\cite{Asai2004}, \cite{Asai2005}, \cite{Asai2007},
\cite{Dursi2007}, \cite{Dursi2008}). In these simulations, the speed of the cold cloud fulfilled the
subsonic and super-Alfv\'enic criteria set out by \cite{Lyutikov2006}. The initial magnetic field
geometries included uniform and tangled fields, and the low-$\beta$ layer formed at the cold front
surface in both cases (see Figure \ref{fig:asai_front}). 

\begin{figure*}
\centering
\includegraphics[width=0.47\textwidth]{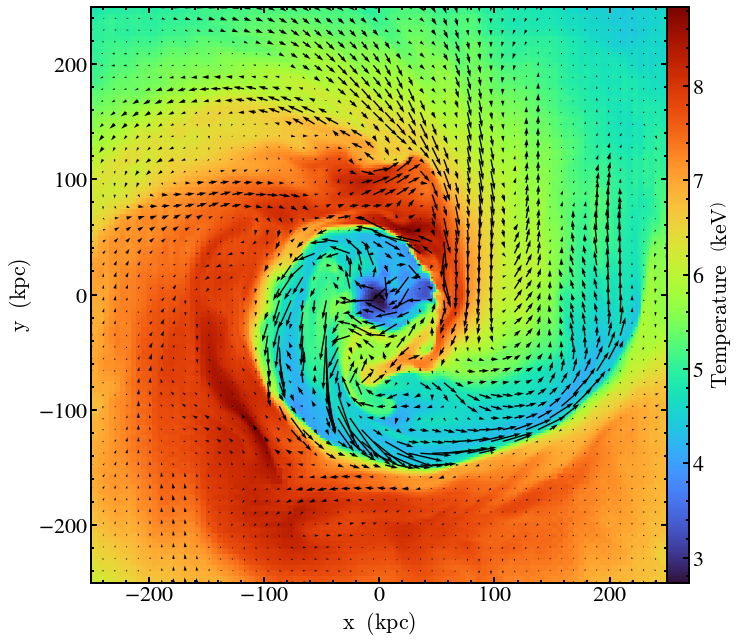}	
\includegraphics[width=0.48\textwidth]{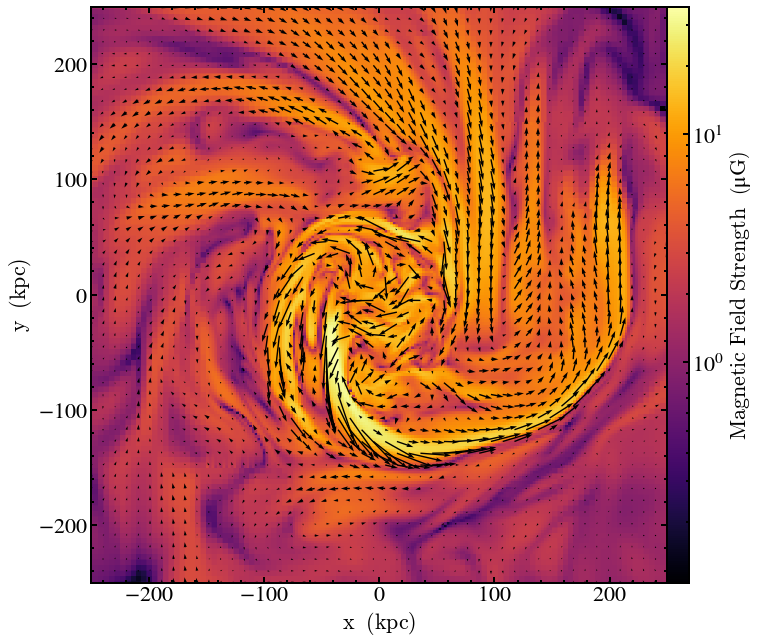}	
\caption{Slices through the gas temperature (left) and magnetic field strength (right) with magnetic field vectors overlaid, showing the amplifcation of magnetic field strength by sloshing motions and alignment of the magnetic field lines with the cold fronts. Figure adapted from \cite{ZuHone2011}.}
\label{fig:sloshing_magnetic}
\end{figure*}
    
Later investigations explored more realistic situations. Simulated idealized cluster mergers with magnetohydrodynamics (MHD) have produced ``merger-remnant'' cold fronts with
magnetic draping layers (see \cite{Takizawa2008} and \cite{Brzycki2019}, though these works did not investigate cold front stability).
Analytical arguments \cite{Keshet2010} have shown that the formation of highly magnetized layers along sloshing cold fronts occurs on the \textit{inside} of the interface, on the colder and denser side, instead of the \textit{outside} of the interface as in the magnetic draping layers mentioned above. The formation of these layers was confirmed by MHD simulations of sloshing cold fronts in a cool-core cluster (see \cite{ZuHone2011} and Figure \ref{fig:sloshing_magnetic}), which also showed that
the magnetic field strengths in these layers were dependent on the initial field strength in the
core region before the sloshing motions begin. For an initial $\beta \sim 100$, the field strengths
in the layers can be quite large, $\beta \lesssim 10$. At these field strengths, KHI is quite
readily suppressed in the simulations. An analysis of several clusters with sloshing cold
fronts indicates a non-thermal pressure fraction under the front surfaces of
$\sim$0.1-0.3, consistent with the expected level of magnetization underneath the front surfaces if
this non-thermal pressure is primarily in the form of magnetic fields \cite{Naor2020}. 

Observations indicate that the magnetic field in clusters may not always be strong enough to
suppress KHI at cold fronts. A deep \textit{Chandra} observation of the nearest cold front in the sky in the Virgo cluster showed that although the most of the cold front had a very narrow width which appeared undisturbed by KHI, the northwestern region appeared to have a width consistent with the presence of KHI eddies on the order of a few kpc \cite{Werner2016a}. Two recent studies compared cold fronts in deep \textit{Chandra}
observations of the Perseus cluster to simulations. The first \cite{Walker2017} identified a ``bay'' feature along a
cold front surface to the SE of the cluster core, which was interpreted as a manifestation of KHI
after analysis of the thermodynamic profiles across the interface. Comparisons of these observations to MHD
simulations \cite{ZuHone2011} with varying initial plasma $\beta$ parameter demonstrated that the
observed level of KHI could be reproduced by the simulation with initial $\beta$ = 200. The second study \cite{Walker2018} showed that another cold front at a larger radius in Perseus was ``split'', and
again showed that this was most consistent with the simulations with initial $\beta$ = 200. 

Another example is the elliptical galaxy NGC~1404, which is falling into the Fornax cluster of
galaxies and has a distinct remnant-core cold front. Deep \textit{Chandra} observations of NGC~1404 \cite{Su2017} showed sub-kpc-scale eddies generated by KHI at the leading edge
of the front. Based on these and the lack of an observed draping layer at the front edge, they
placed a constraint that any magnetic field draping layer must be no stronger than $\sim$5~$\mu$G. 

\begin{figure*}
\centering
\includegraphics[width=0.95\textwidth]{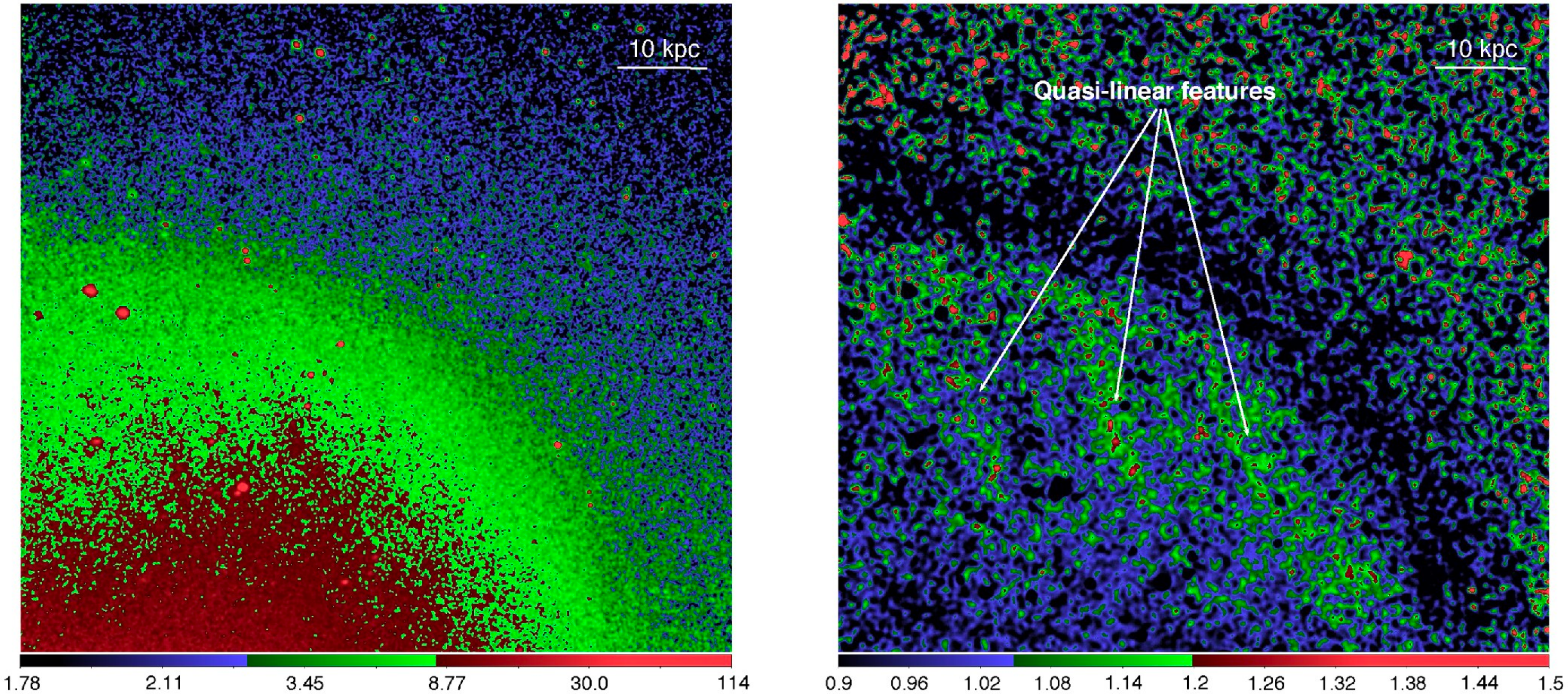}	
\caption{A deep \textit{Chandra} observation of the NW cold front in the Virgo cluster, reproduced from \cite{Werner2016a}. {\it Left}: Raw counts image. {\it Right}: Residual counts image, obtained by dividing the image on the left by an azimuthally symmetric model, showing the cold front edge and the puzzling linear features underneath it, which may be caused by amplified magnetic fields.}
\label{fig:virgo_cold_front}
\end{figure*}
    
It has also been suggested that magnetic field amplification by mergers could be responsible for the appearance of linear features in some X-ray observations of clusters, such as the peculiar surface brightness structures underneath the cold front in the Virgo cluster \cite{Werner2016a}, the ``feathers'' in the Perseus cluster \cite{Ichinohe2019}, or the narrow ``channels'' in other clusters, such as Abell 520 \cite{Wang2016} and Abell 2142 \cite{Wang2018}. In these cases, it is theorized that an initially tangled magnetic field has been stretched along a particular direction by gas motions, with areas of initially stronger field producing dips in surface brightness along that direction. This occurs because while the magnetic pressure increases, the total pressure remains the same and thus the thermal pressure and the density in the feature decreases, thus also decreasing the X-ray surface brightness. These narrow, low-surface brightness features are surrounded on either side by areas of weaker field and thus brighter emission. Alternatively, a similar phenomenon may present wider strips of lower surface brightness alternated by narrower, brighter areas such as in Virgo, see Figure \ref{fig:virgo_cold_front}).

\subsection{Thermal Conduction}\label{sec:conduction}

\begin{figure}
    \centering
    \includegraphics[width=0.95\textwidth]{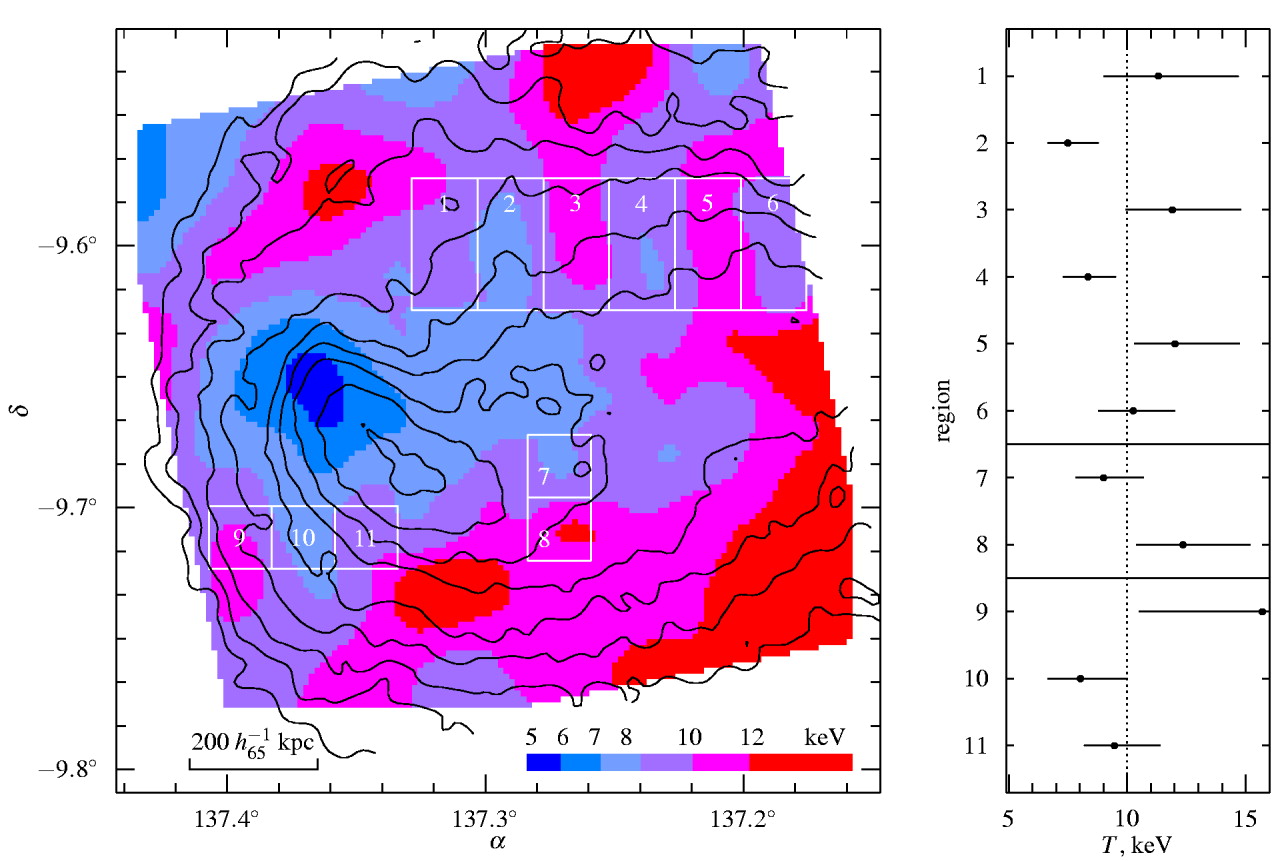}
    \caption{\textit{Chandra} temperature map of the merging cluster Abell~754 with X-ray surface brightness contours overlaid. Temperature fits for regions shown in the white rectangles are shown in the right panel (with 90\% errors). Reproduced from \cite{Markevitch2003}.}
    \label{fig:A754}
\end{figure}

Cluster mergers will drive temperature differences in the ICM by bulk motions and turbulence which should be quickly smoothed out by thermal conduction \cite{Dolag2004}. The \textit{Chandra} observations of Abell~754 showed its ICM has a patchy temperature structure, which was used to constrain the thermal conduction at an order of magnitude under the Spitzer value (see Figure~\ref{fig:A754} and \cite{Markevitch2003}). The most powerful constraints on thermal conduction potentially come from cold fronts, which have been explored by a number of simulations. Hydrodynamic simulations of a merger-remnant cold front similar to the one in Abell 3667 with isotropic thermal conduction used the width of the cold front in the simulations to
determine the level of conductivity consistent with the observed width \cite{Xiang2007}. The \cite{Asai2004},
\cite{Asai2005}, \cite{Asai2007} simulations mentioned above showed that magnetic draping layers
were able to significantly suppress anisotropic thermal conduction, while isotropic conduction in
the same simulations smoothed out cold fronts on fast timescales (see also \cite{Suzuki2013}). Simulations were carried out by \cite{ZuHone2013a} that included anisotropic thermal conduction in the same setups in \cite{ZuHone2011} of sloshing cold fronts with magnetic fields, with varying levels
of conductivity along the field lines. In this case, it was found that while cold fronts still formed,
if heat conducted along the field lines at the Spitzer level, the temperature and density gradients
across the front were greatly reduced, effectively making the fronts unobservable in X-rays. How is
this possible, given the existence of the magnetic layer tangential to the front surface? Unlike the
very simplified simulations of an idealized cold front in a uniform medium (\cite{Asai2004},
\cite{Asai2005}, \cite{Asai2007}), the magnetic field does not form a layer which completely
surrounds the cold front. In the sloshing simulations, the cold gas inside the front was connected
via magnetic field lines to hotter gas from which heat could conduct to increase the former's
temperature. Additionally, imperfections in the magnetic layer due to KHI or turbulence could
establish connections between the two sides of the front, also allowing heat to conduct between
them (see also \cite{ZuHone2015a} for the same results from a different cluster setup). These results suggest that the thermal conductivity of the ICM, even
parallel to the magnetic field lines, may be at least an order of magnitude less than the Spitzer
value. The deep observations of the Virgo cold front mentioned above \cite{Werner2016a}, with the
very narrow width of the interface on the order of a few Coulomb mean free paths, support this
conclusion. 

The suppression of conduction by a magnetic field is also implied by the existence of the bright X-ray tails downstream ram-pressure-stripped early-type galaxies in the ICM \cite{Eckert2014,Eckert2017a}. Simulations of ram-pressure-stripped early-type galaxies \cite{Vijayaraghavan2017a} showed that the formation of long-lived X-ray \textit{coronae} and ram-pressure-stripped tails are suppressed when conduction is isotropic and not suppressed by a magnetic field; the follow-up study \cite{Vijayaraghavan2017b}, which included magnetic fields and anisotropic thermal conduction, showed that the observed features can survive in this circumstance, though it depended on the magnetic field geometry. 

\subsection{Viscosity}\label{sec:viscosity}

Constraining the viscosity of the ICM via cold fronts has proven to be more challenging. The first
reason for this is straightforward; the main observable characteristic used to constrain viscosity
is the appearance of KHI along front surfaces, which can be suppressed by viscosity, magnetic
fields, or a combination of both. It may also be the case that there has simply not been enough time for KHI to develop. 

In practice, simulations have often focused on either the effects of magnetic fields or viscosity, and not modeled them together. In hydrodynamic simulations, the anisotropy of viscosity has usually been
accounted for by adding a multiplicative suppression factor to the Spitzer viscosity. MHD simulations have incorporated Braginskii (anisotropic) viscosity.

\begin{figure*}
\centering
\includegraphics[width=0.95\textwidth]{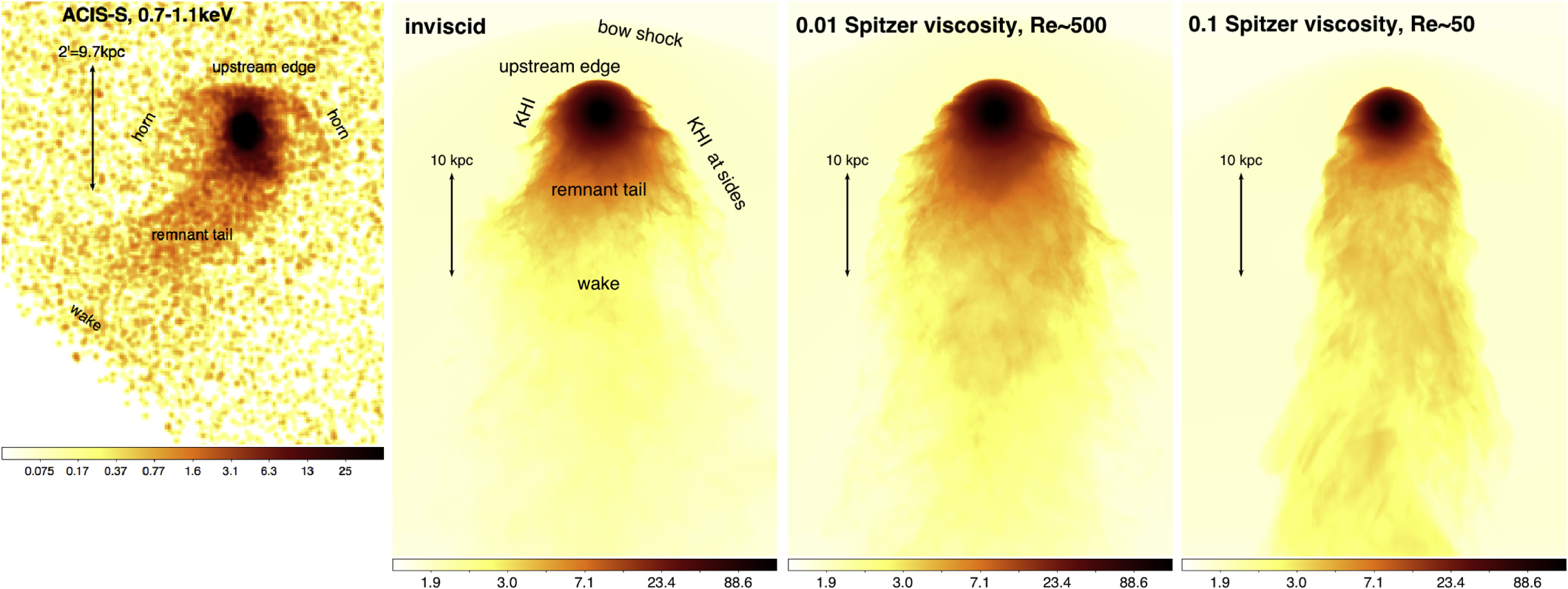}	
\caption{The effect of viscosity on ram-pressure-stripped tails of galaxies falling through the ICM, reproduced from \cite{Roediger2015b}. {\it Left panel}: \textit{Chandra} image of the gas-stripped elliptical galaxy M89 in Virgo, with features of the cluster atmosphere and the remnant tail labeled. {\it Remaining panels}: Simulations of the M89 system with varying levels of viscosity. As viscosity increases, the ``horn'' and KHI-driven features are suppressed, and the remnant tail becomes much longer. High viscosities (rightmost panel) are inconsistent with the observations.}
\label{fig:M89}
\end{figure*}

The most advanced simulations of the formation of a remnant-core cold front, a ram-pressure-stripped tail, and their subsequent
evolution with isotropic viscosity are those of \cite{Roediger2015a,Roediger2015b}, who simulated
the infall of the elliptical galaxy M89 into the Virgo cluster and followed the progressive gas
stripping process. A remnant-core cold front forms, stretching to the sides of the galaxy's hot
atmosphere. Behind the galaxy a ``remnant-tail'' forms. They showed that the shape and width of the
cold front and the length of the remnant-tail is strongly dependent on the viscosity--even a small
viscosity of 0.1 the Spitzer value appears inconsistent with the \textit{Chandra} observations of
M89 (see Figure \ref{fig:M89}). Similar conclusions were drawn by \cite{Su2017} from the
aforementioned observations of NGC~1404, who put an upper limit of 5\% Spitzer on the isotropic
viscosity of the ICM based on the KHI at the leading edge of the front. They also observed mixing
between the hot cluster gas and the cooler galaxy gas in the downstream stripped tail, providing
further evidence of a low-viscosity plasma. The inviscid simulations of the NGC~1404/Fornax system
by \cite{Sheardown2018} appear very consistent with the \textit{Chandra} observations in these
respects, supporting the assumption of a low viscosity.

The only investigation so far of remnant-core cold front evolution
with Braginskii has been carried out by \cite{Suzuki2013}, using a setup very similar to \cite{Asai2004,Asai2005,Asai2007}. The
magnetic fields in these simulation were made very weak so that any suppression of KHI
would be from viscosity alone. Different field line orientations were experimented with, and the simulations showed
that Braginskii viscosity is not as efficient as an isotropic viscosity at suppressing KHI because
its damping effect of viscosity is reduced by the dependence on the field line direction. 

With regard to sloshing cold fronts, the first systematic study of the effect of isotropic viscosity
was carried out by \cite{Roediger2013}. They showed that an inviscid vs. a viscous ICM could be
distinguished by characteristic cold front features; in the absence of viscosity the surface
brightness edge appears ragged and profiles across the surface exhibit a ``multi-step'' structure.
These features were absent in an otherwise identical simulation with 0.1 Spitzer viscosity. As mentioned above, the deep
\textit{Chandra} observations of the Virgo cold front in \cite{Werner2016a} exhibited a non-zero width in the northwestern region consistent with KHI eddies. This was used to estimate that the effective viscosity in the Virgo ICM is suppressed by at least an order of magnitude below the Spitzer value. The Virgo cluster was also simulated by \cite{ZuHone2015a}, using the same setup as
\cite{Roediger2013}, but including weak magnetic fields and compared anisotropic Braginskii viscosity
to isotropic Spitzer viscosity. At least from the perspective of KHI at cold front surfaces, Braginskii viscosity had similar effects as isotropic viscosity which operated at 0.1 Spitzer. 

Finally, developments in high-$\beta$ plasma studies indicate that interactions between the
electrons and ions with the magnetic field at Larmor-radius scales may substantially modify the
viscosity and the thermal conductivity of the ICM; for more details we refer the reader to Chapter ``Plasma Physics of the ICM/IGrM''.

\subsection{Electron-Ion Equilibration at Cluster Shocks}\label{sec:shock_ei_eq}

\begin{figure*}
\centering
	\includegraphics[width=0.32\textwidth]{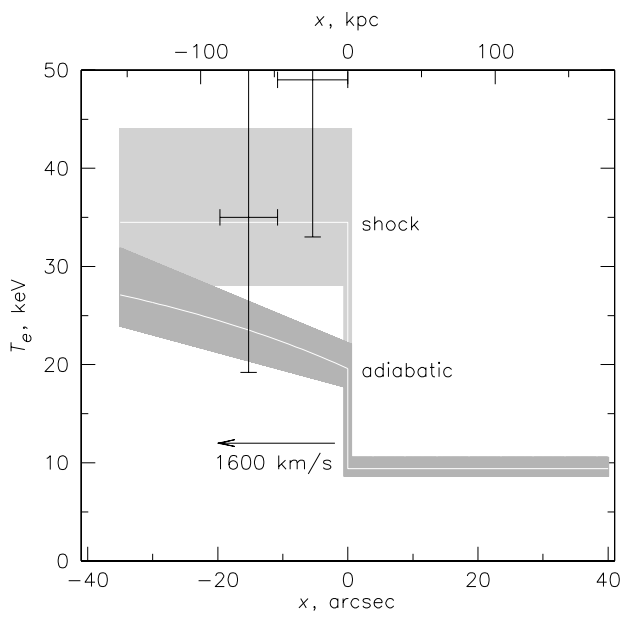}	
	\includegraphics[width=0.3\textwidth]{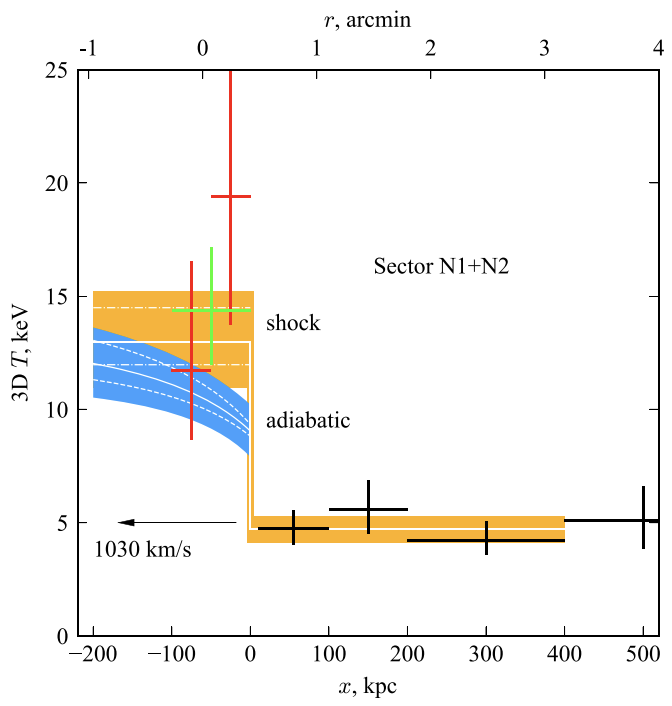}	
	\includegraphics[width=0.3\textwidth]{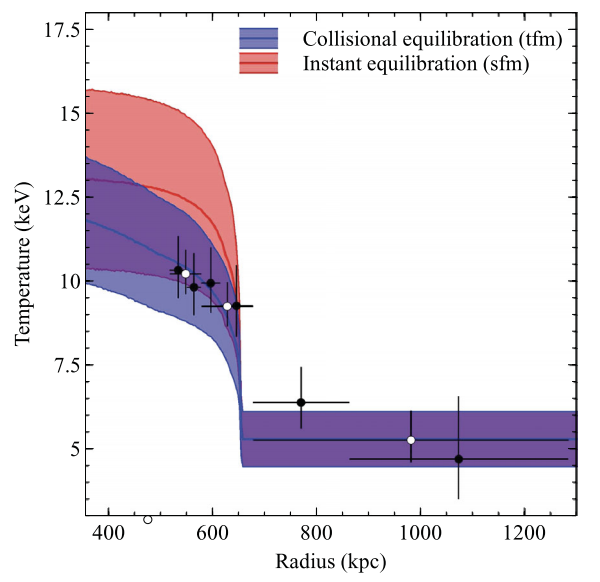}
    \caption{The measured electron temperature profiles compared with model predictions for instant equilibration (shock) and adiabatic compression followed by collisional equilibration (adiabatic). From {\it left} to {\it right} are for the cases of 1E 0657-56 \cite{Markevitch2002}, Abell~520 \cite{Wang2016}, and Abell~2146 \cite{Russell2012}, respectively. Unlike 1E 0657-56 and Abell~520, the profiles of Abell~2146 are projected along the line of sight. }
    \label{fig:ei}
\end{figure*}

Current major X-ray telescopes such as \textit{Chandra}, \textit{XMM-Newton}, and \textit{NuSTAR} can only directly measure the temperature of electrons in the ICM through a continuum spectrum from bremsstrahlung. The X-ray microcalorimeter spectrometer onboard \textit{Hitomi} provided the first measurement of the ion temperature in the ICM using the broadening of the well-resolved emission lines in the Perseus Cluster \cite{Hitomi2018}, which was in agreement with the electron temperature (though with large uncertainties). Despite the fact that the \textit{equilibration timescale} between electrons and ions is much longer than those for electrons and for the ions separately, it is commonly assumed that ions and electrons share the same temperature. This is very likely to be the case in environments such as cluster cores, where the densities are large enough that the timescales are short. However, for less dense environments, where interactions between electrons and ions which would bring them to a common temperature are less frequent, this assumption (and potential deviations from it) can have a profound impact on cluster astrophysics and cosmology, such as the computation of hydrostatic mass estimates and the entropy profiles of cluster outskirts (see Chapters ``Thermodynamical Profiles of Clusters and Groups, and their Evolution'' and ``Cluster Outskirts and their Connection to the Cosmic Web'' for details). 

The passage of a shock front is likely to affect electrons and ions differently. For the collisional plasma, shock passage may heat ions immediately, while electrons with a thermal velocity much higher than the shock velocity are compressed adiabatically to a temperature lower than the ion temperature:
\begin{equation}
T_{\rm e,2}= T_{\rm e,1}\left(\frac{\rho_2}{\rho_1}\right)^{\Gamma-1}
\label{eq:ad} 
\end{equation}
where $\Gamma$ is the ratio of specific heats for a monatomic gas. Electrons and ions will eventually reach thermal equilibrium via Coulomb collisions over an electron-ion equilibration timescale of 
\begin{equation}
    t_{\rm ei}=6.2\times10^8 ~{\rm yr}\left(\frac{T_{\rm e}}{10^8\,{\rm K}}\right)^{3/2}\left(\frac{n_{\rm e}}{10^{-3}{\rm cm^{-3}}}\right)^{-1},
\label{eq:tei} 
\end{equation}
where $T_{\rm e}$ and $n_{\rm e}$
are the electron temperature and density, respectively \citep{Spitzer1962}.

Since the total kinetic energy density $U=\frac{3}{2}k(n_{\rm i}T_{\rm i}+n_{\rm e}T_{\rm e})$ is conserved, the mean gas temperature given by
\begin{equation}
T_{\rm gas}=\frac{(n_{\rm i}T_{\rm i}+n_{\rm e}T_{\rm e})}{n_{\rm i}+n_{\rm e}}=\frac{T_{\rm i}+1.21T_{\rm e}}{2.21}
\label{eq:Tgas} 
\end{equation}
(assuming $n_{\rm e} = 1.21n_{\rm i}$) should be constant with time and can be calculated from the Rankine-Hugoniot shock jump conditions, which provides a means to measure the immediate post-shock ion temperature. 
The electron temperature subsequently increases while the ion temperature decreases due to the Coulomb collision at a rate of 
\begin{equation}
    \frac{{\rm d}T_{\rm e}}{{\rm d}t}=-(n_{\rm i}/n_{\rm e})\frac{{\rm d}T_{\rm i}}{{\rm d}t}=\frac{T_{\rm i}-T_{\rm e}}{t_{\rm ei}}
\label{eq:rate} 
\end{equation} 
Rearranging this equation to obtain 
\begin{equation}
    \frac{t_{\rm ei}}{T_{\rm i}-T_{\rm e}}{\rm d}T_{\rm e}={\rm d}t
\label{eq:dt} 
\end{equation}
By integrating Equation~\ref{eq:dt} and using Equations~\ref{eq:tei} and \ref{eq:Tgas}, one can obtain $T_{\rm e}$ as a function of time, which can be transformed to $T_{\rm e}$ as a function of distance behind the shock front (see e.g. \cite{Fox1997, Ettori1998}). However, the actual equilibration timescale may be shorter, if (for example) microscale plasma processes play a role in mediating interactions between electrons and ions (see Chapter ``Plasma Physics of the ICM/IGrM'').

If the actual electron heating timescale is shorter than the equilibrium time $t_{\rm ei}$, the measured electron temperature would be higher than that expected from the Coulomb-based collisional model. In the limit that the equilibration between electrons and ions is instantaneous, the post-shock electron temperature would be the same as the ion temperature, which can be obtained from the shock jump conditions. 
To test these scenarios, it is critical to measure the temperature profile behind the shock front on a scale of the product of $t_{\rm ei}$ and shock velocity, which is $\sim$\,tens of kiloparsecs and can be resolved by \textit{Chandra} for strong shocks in nearby massive clusters. The time evolution of the electron temperature downstream along a merger shock has been first measured in the 1E 0657-56 \cite{Markevitch2006}. As shown in Figure~\ref{fig:ei}-left, the electron temperature exceeds the expectation from the collisional model and conforms with the instant shock heating of the electrons. 
Note that the measured post-shock temperature is $\sim30$\,keV while the energy response of \textit{Chandra}-ACIS ranges between 0.1\,keV and $10$\,keV, which casts some doubt on this result. Presumably, a hard X-ray telescope such as \textit{NuSTAR} would be more sensitive to such temperatures. However, the modest spatial resolution of \textit{NuSTAR} makes the inclusion of the lower energy photons surrounding the shock front unavoidable, which can bias the temperature low.

The same test has been performed with the bow shock in Abell~520 where the post-shock temperature is 15\,keV, more in line with \textit{Chandra}'s capabilities. As shown in Figure~\ref{fig:ei}-middle, 
its electron temperature immediately behind the bow shock is higher than expected for the adiabatic compression scenario, suggesting that a faster equilibration mechanism is at work in the magnetized ICM \cite{Wang2018}. However, the gas distribution of Abell~520 
contains remarkably rich substructures,
resembling a ``train wreck site". The detailed structure of the gas in Abell~520 could complicate the analysis of the downstream temperature profile.     
Perhaps a more ideal laboratory is presented by the bow shock in Abell~2146 \cite{Russell2012}. Its post-shock temperature is at $\sim10$\,keV and its collision geometry is relatively simple. As shown in Figure~\ref{fig:ei}-right, the temperature profile behind the bow shock favors the collisional equilibration model.  
The discrepancy of the above studies is possibly due to the uncertainties in merging geometry such as the curvature of the shock front along the line of sight, or it may reflect more detailed physics in the different situations. 
Expanding the cluster shock sample as well as performing simulations specifically tailored to each cluster may be needed to draw any firm conclusion on the electron heating mechanism. 

\section{Merging Clusters and Cosmic Rays: Observable Signatures in the Radio and X-ray Bands}\label{sec:cosmic_rays}
    
A small portion of the baryonic material in clusters is in the form of CRs. Cosmic-ray protons (CRp)
and electrons (CRe) are produced by the acceleration of particles by supernovae and AGN in galaxies, as well as turbulence and shocks in the ICM. After being injected into the ICM, CRs advect with it by being constrained by the magnetic field which is frozen into the fluid, and stream and diffuse down the field lines \cite{Ensslin2011}. 

\begin{figure}
\centering
\includegraphics[width=0.47\textwidth]{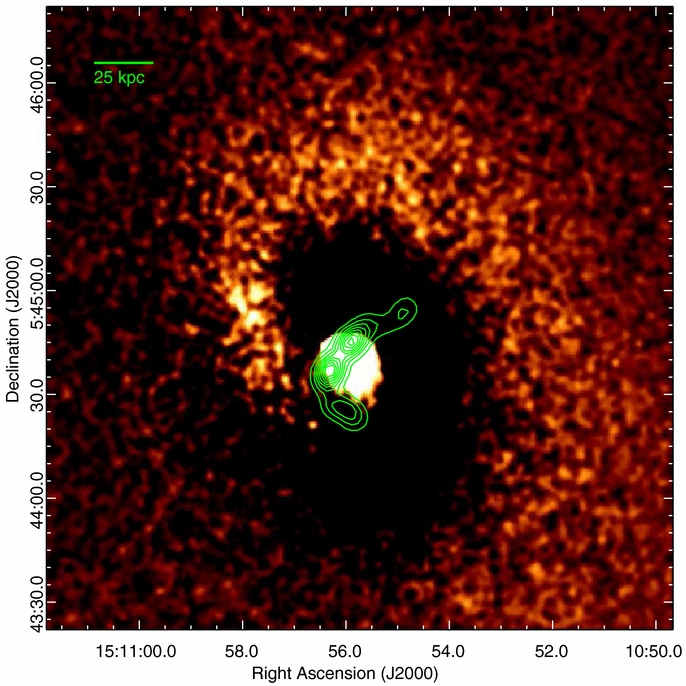}	
\includegraphics[width=0.47\textwidth]{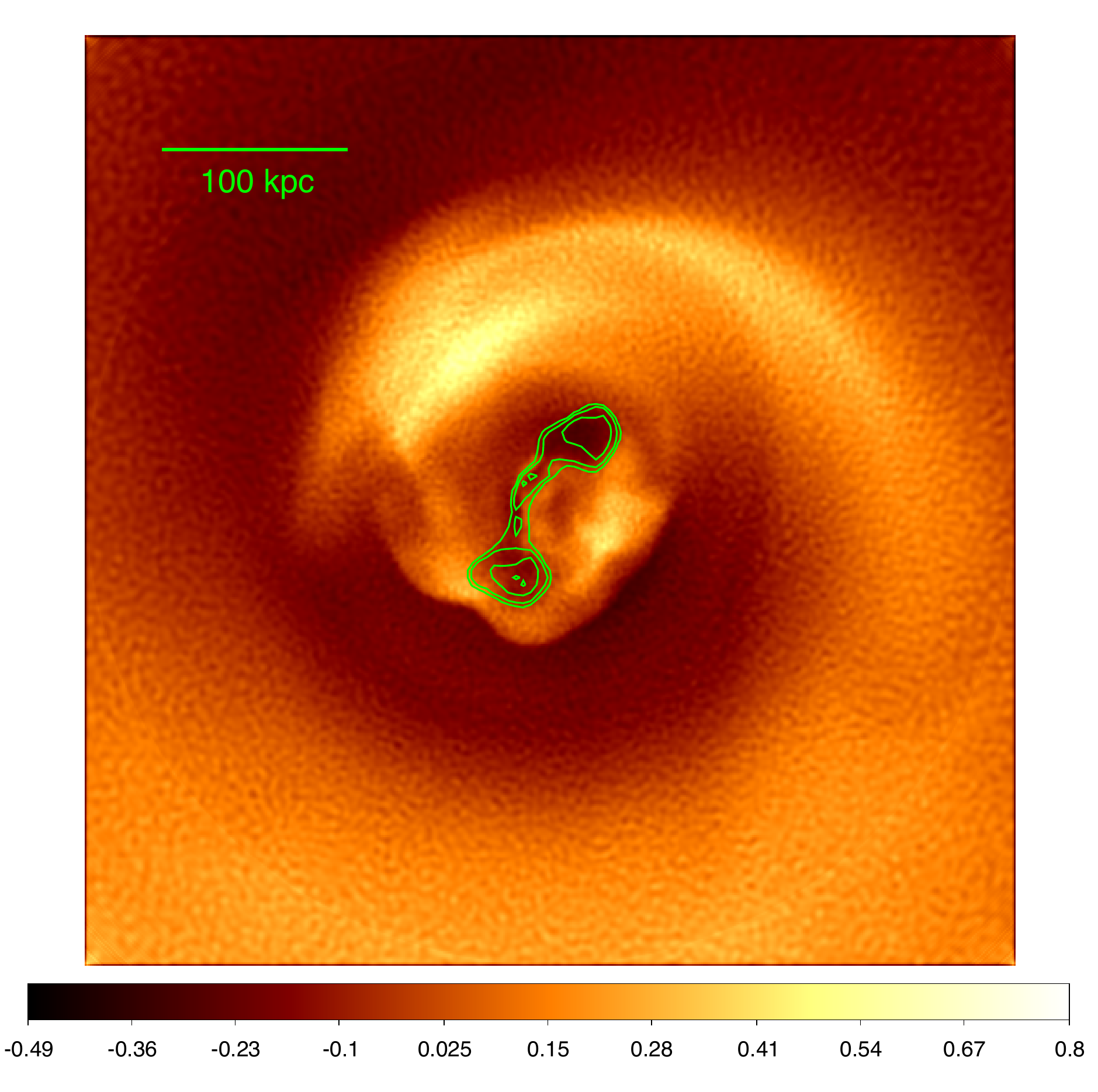}	
\caption{{\it Left}: Residual X-ray surface brightness image of the core of the cluster Abell 2029, with 1.4~GHz contours from the central AGN overlaid, from \cite{PaternoMahler2013}. The sloshing motions of the ICM are bending the radio jets. {\it Right}: Residual X-ray surface brightness image from a simulation of a sloshing cluster core with contours of the CR energy density from the central AGN in the simulation overlaid, showing the bending of jets due to sloshing. Adapted from \cite{Fabian2022}.}
\label{fig:wat}
\end{figure}

CRe with $\gamma \sim 10^3-10^4$ (where $\gamma = 1/\sqrt{1-v^2/c^2}$ is the relativistic Lorentz factor) emit \textit{radio synchrotron radiation} in the $\sim\mu$G magnetic field
of the ICM. When injected by AGN jets, they produce associated \textit{radio lobes} which can encounter merger-driven ICM bulk motions and be pushed away from their initial trajectory, producing \textit{wide-angle tails (WATs)}. A famous example of this is in the cluster Abell 2029, where the sloshing motions have produced a WAT (see \cite{Clarke2004,PaternoMahler2013} and Figure \ref{fig:wat}). Recent simulation works have shown that the bulk motions expected in cluster cores can easily produce these features \cite{Mendygral2012,Fabian2022}.

There are other, more diffuse, and larger radio features in clusters which are produced by CRe. 
However, these electrons have short cooling times, and will take much longer to diffuse away from their sites of origin (supernova, SNe, etc.) before they are no longer able to shine in the radio.  Therefore, they must be accelerated to
these energies by interactions with the ICM, or produced as secondary particles from \textit{``hadronic''
interactions} between CRp and thermal protons in the ICM \cite{Keshet2011}. Given either or both of these possibilities at play, features in the X-ray band are likely to be spatially correlated with features produced by CRe in the radio band. 

ICM turbulence driven by mergers can reaccelerate lower-energy CRe \cite{Brunetti2007} up to $\gamma
\sim 10^4$ where they will emit synchrotron radiation in the radio band. Such is thought to be the
origin of \textit{``radio halos''}, Mpc-scale diffuse and unpolarized emission which is often seen in
clusters undergoing major mergers. Such emission should correlate with signatures of turbulent
motions in the ICM, whether indirectly via surface brightness and/or pressure fluctuations \cite{Eckert2017b} or
directly via measurements of gas motions from microcalorimeter instruments on future X-ray
observatories such as \textit{XRISM}, \textit{Athena}, and the proposed mission concept \textit{Lynx}. The generation of a radio halo was investigated in idealized simulations of a major merger \cite{Donnert2013} , where the CRe were reaccelerated by cluster turbulence. The appearance of the radio halo emission in those simulations was strongly dependent on the merger stage, which can be constrained by comparing X-ray observations to simulations. 

\textit{``Radio mini-halos''} are the smaller-scale cousins of radio halos (a few hundred kpc wide), and
usually appear in more relaxed systems which show signatures of sloshing motions. Often, radio
mini-halos have boundaries which correspond to the locations of sloshing cold front edges
\cite{Mazzotta2008,Giacintucci2014,Giacintucci2021}, suggesting that the turbulence driven by the
sloshing motions and/or the magnetic field amplification which occurs underneath the cold fronts
play a role in the appearance of these features \cite{ZuHone2013b}. Alternatively, it is possible
that radio mini-halos can be produced via hadronic interactions \cite{Pfrommer2004a,Pfrommer2004b,Ignesti2020}, which would then have their association with the sloshing cold fronts determined by the magnetic field \cite{ZuHone2015b} or by a combination of diffusion and advection of CRp generated from the central AGN within the sloshing cold fronts \cite{ZuHone2021b}. Which of these two physical processes plays the more important role in producing mini-halos can be determined in the future by constraints on the turbulent motions in these systems by microcalorimeters on future X-ray missions.

\begin{figure*}
\centering
	\includegraphics[width=0.4\textwidth]{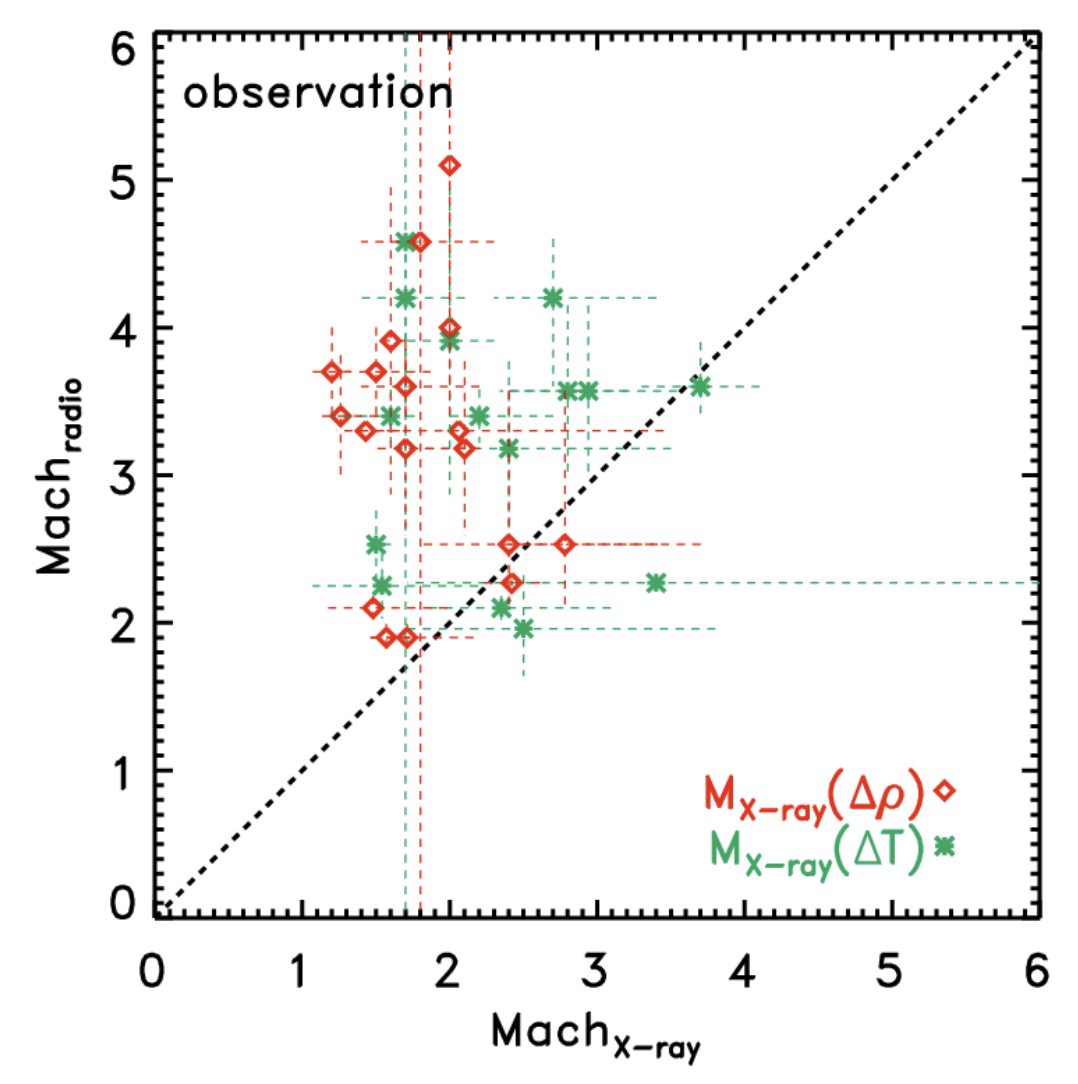}	
		\includegraphics[width=0.51\textwidth]{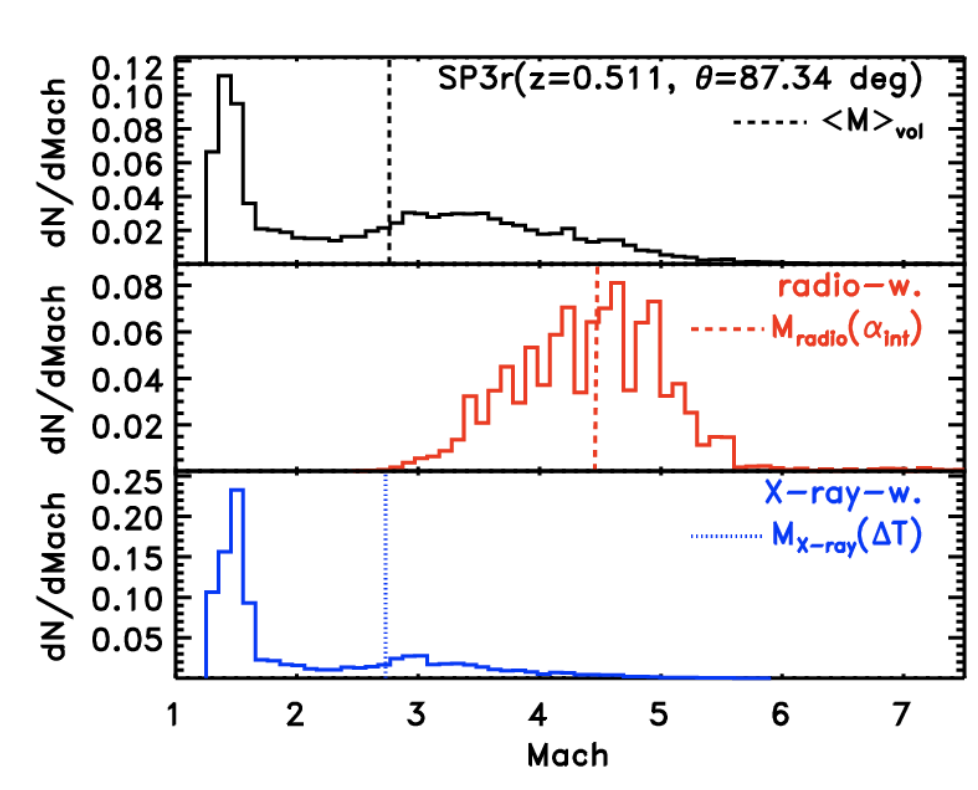}	
    \caption{{\it Left}: Discrepancies of Mach numbers determined through X-ray and radio observations taken from \cite{Wittor2021}. Note that the Mach numbers derived with X-ray temperature and X-ray density are not consistent, which may be caused by their different dependencies on the viewing angle. {\it Right}: 3D Mach number distributions for the relics. The three panels
give the volume (black, top), radio (red, middle) and X-ray (blue, bottom) weighted Mach number distribution of the shock measured
in 3D taken from \cite{Wittor2021}.}
    \label{fig:wittor}
\end{figure*}

\textit{``Radio relics''} are diffuse, extended, and polarized radio sources observed at large radii in major cluster mergers. These features are often associated with bow shocks as seen in the X-ray. This correspondence gave rise to the \textit{diffusive shock acceleration (DSA)} theory \cite{Drury1983,Blandford1987} that the emitting CRe arise from acceleration of electrons from the thermal pool up to the requisite relativistic energies by crossing a shock front multiple times. The end result is a power-law energy distribution of CRe. Assuming this model, radio relics could therefore provide an independent measurement of the shock Mach number from the integrated radio spectrum index of a power law:
\begin{equation}
I_{\nu}\propto\nu^{-\alpha_{\rm int}}, ~~~\alpha_{\rm int}=\frac{{\cal M}^2+1}{{\cal M}^2-1}
\end{equation}
Mach numbers derived from X-ray observations are systematically lower than those from radio analysis for the same cluster shocks \citep{Wittor2021}, as shown in Figure~\ref{fig:wittor}. X-ray analyses are more vulnerable to projection effects which can reduce the sharpness of the edges (and lead to a lower measured Mach number). On the other hand, simulations show that radio detections are biased towards strong shocks while X-ray detections are more representative of the average shock distributions (Figure~\ref{fig:wittor}).

The DSA scenario is not without its difficulties. In some cases, extrapolating the power-law spectrum of the CRe down to the energies of the electrons in the thermal ICM yields comparable energy densities, implying an implausibly high acceleration efficiency assuming the DSA model \cite{Macario2011}. Also, the predicted $\gamma$-ray emission by the interaction of CRp with thermal protons in these shocks in the DSA model is higher than the upper limits in clusters observed by the \textit{Fermi} observatory \cite{Vazza2014,Vazza2015}. Finally, in some cases the expected acceleration efficiency from the standard DSA model is not adequate to produce the observed radio luminosity \cite{Botteon2020}.

For these reasons, it is much more likely that a population of CRe with smaller energies ($\gamma \sim 10^2$), which have longer cooling times, are re-accelerated by the shock fronts \cite{Markevitch2005,Kang2011,Kang2016,Pinzke2013}. However, not all radio relics have easily identified shock counterparts, which may indicate they are too faint to be seen with current instruments, or that the shapes of radio relics in some cases may have more to do with the shape of the underlying CRe distribution as they are injected by AGN \cite{ZuHone2021a,ZuHone2021b,Vazza2021}. 

The same CRe which produce synchrotron radio emission in clusters (regardless of their origin)
should also produce very high-energy X-rays from \textit{inverse-Compton (IC) scattering} off of cosmic microwave background photons. A power-law distribution of CRe should produce a
power-law spectrum of X-rays via IC scattering. In principle, a determination of the ratio of the radio synchrotron emission to the IC emission from the same electrons would permit an independent estimation of the volume-averaged magnetic field strength and from that also derive the energy density of the CRe.

In realistic cluster scenarios, this emission would
be several orders of magnitude fainter than the thermal emission from the ICM at energies less than
several tens of keV. This places confident IC detections outside of the bandpass of past and current
X-ray observatories, with the exception of \textit{Suzaku} and \textit{NuSTAR}, as well as the
$\gamma$-ray instruments \textit{INTEGRAL} and \textit{Swift/BAT}. A high-energy tail of X-ray
emission in the Ophiuchus cluster may be produced by IC scattering from the same electrons that
produce the mini-halo in that system \cite{Eckert2008,Fujita2008,Nevalainen2009,Murgia2010}, though
this claim has been disputed \cite{Colafrancesco2009}. Otherwise, no clear detections of IC X-ray
emission in clusters have so far been made; only upper limits exist
\cite{Wik2011,Wik2012,Wik2014,Cova2019,RojasBolivar2021}. The situation is complicated by the fact
that in many of the same clusters where one would expect turbulent or shock (re)acceleration of CRe
which produce IC emission, there is also the presence of very hot shocked plasma which will emit at
the same high energies of over 10 to a few 10s of keV; in most cases, detections of excess emission at high energies have been fit
equally well or better by a secondary thermal component rather than IC emission. 

\section{Conclusions}

Mergers between clusters of galaxies with other clusters, groups, and galaxies are the most powerful collisions in the current universe. X-ray observations of merging clusters have revealed that their ICM is very dynamic, and the observed features have been confirmed by numerical simulations of cluster mergers. 

We close this chapter with a summary, and pointers to the new discoveries that future X-ray observatories can make:

\begin{itemize}

\item {\bf Cold fronts:} Cold fronts are contact discontinuities in the ICM formed when bulk motions driven by mergers bring gases of different entropies into contact. Given that gas motions of some kind are nearly ubiquitous in clusters, cold fronts are found in the vast majority of X-ray observations of these systems. Formed by major mergers or gas sloshing set off by minor mergers, these features are typically easily observed due to their association with bright, dense gas. Due to the shear motions associated with their formation and subsequent propagation, cold fronts should be sites of enhanced magnetic field, as predicted by simulations and hinted at by observations. Plasma depletion layers associated with cold fronts, if they exist, will be more easily detected with future X-ray observatories with large effective area. 

\item {\bf Shock fronts:} 
Merger shocks are driven by the supersonic bulk motion of cluster gas. Shock fronts are X-ray brightness edges associated with a temperature elevation on the brighter side, creating a pressure jump in the ICM. Merger shocks are a primary heating source for the hot and energetic cluster gas. They are also ideal laboratories for studying the processes which drive electron-ion equilibrium and the acceleration of relativistic particles. 
Confirmed shock fronts are relatively rare, and are most likely to be detected during cluster mergers shortly after the first core passage, when they are near the center of the merging system and there are sufficient X-ray photons to constrain the temperature. Their detections are also sensitive to the merger geometry relative to the plane of the sky. Nevertheless, pre-merger shocks at cluster outskirts have also been expected and detected. Numerous cluster shocks at various merger phases remain to be explored for future studies in both observations and simulations. 

\item {\bf Ram-pressure-stripped and slingshot tails:}
Gaseous tails without a stellar counterpart provide unambiguous evidence for the motion of galaxies or groups through the ICM.  In most cases, the tails are formed from ram pressure stripping of the ISM, trailing behind the stellar body. For late-type galaxies, the stripped tail can be multi-phase due to the mixing of the cold ISM and hot ICM. For early-type galaxies, the morphology and thermal structure of their stripped X-ray tail can be fundamentally modified by the ICM plasma conditions such as thermal conductivity, viscosity, and magnetic field structure.  
Recently, a new type of X-ray tails has been identified for early-type galaxies or groups during pericenter passage — slingshot tails. They are misaligned with the direction of motion, likely caused by tidal forces rather than ram pressure stripping.

\item {\bf Velocity measurements:} The next frontier of X-ray observations of clusters will be defined by high-resolution spectroscopy, which will enable the measurement of Doppler shifting and broadening of emission lines due to gas motions. Mergers will drive bulk motions (which will be primarily measured through line shifts) and turbulence (which will be primarily measured through line widths). Based on the \textit{Hitomi} observations of Perseus, it is to be expected that future observations of merging clusters may yield some surprising results and will need to be interpreted through detailed simulations. With the advent of high-angular resolution microcalorimeters onboard \textit{Athena} and perhaps \textit{Lynx}, detailed velocity maps will be made. These will be compared to estimates from surface brightness fluctuations and used to compute the velocity power spectra of the gas. Additional insights into the dynamics of the different components of a cluster merger including gas, stars, and DM will also be obtained via synergies of this technique with spectroscopic-derived galaxy velocities and resolved kinetic Sunyaev-Zeldovich measurements of the ICM velocity\footnote{The Sunyaev-Zeldovich (SZ) effect is the distortion of the cosmic microwave background spectrum by thermal electrons in the ICM. The ``kinetic'' SZ effect arises from Doppler shifting of the cosmic microwave background photons due to bulk motions in the ICM.}. 

\item {\bf Plasma physics:} The existing X-ray observations of clusters have provided some constraints on the detailed properties of the ICM plasma. \textit{Chandra} is especially powerful in this sense, because it can probe small length scales thanks to its sub-arcsecond angular resolution. Deep exposures of cold fronts and possible X-ray channels have provided some initial indications of the magnetic field strength in these regions of clusters. The viscosity and thermal conduction of the ICM are as yet unknown to great precision; however, the available observations of cold fronts and ram-pressure-stripped tails appear to indicate that the ICM is at most mildly viscous and that conduction is likely strongly suppressed. The electron-ion equilibration timescale, another probe of ICM plasma physics, is not yet well-constrained based on the few observations of merger shocks capable of providing such measurements. Advances in these areas will be made with future X-ray observatories with high angular-resolution and large effective area which can resolve the relevant features with much greater precision. Additionally, high-angular resolution maps of turbulent gas velocities using microcalorimeters on future missions can also provide constraints on viscosity.  

\item {\bf High-energy particles and synergies with radio observations:} The relativistic CRe in clusters shine most brightly in the radio band, but are nearly always associated with features in the X-ray due to their interactions with the ICM. The large and diffuse radio halos are believed to be produced by turbulent re-acceleration of CRe, and future instruments will be able to more accurately constrain this gas turbulence via measurements of surface brightness fluctuations or actual measurements of the turbulent velocity with microcalorimeters. The same is true of the smaller radio mini-halos, which will also provide a test of the contribution to the mini-halo emission from hadronic models. An X-ray observatory with a larger effective area will detect more shock fronts in the faint regions of merging clusters, including those with radio relic counterparts. These will be used to constrain the model space for acceleration of CRe by these shocks. Finally, future X-ray or $\gamma$-ray observatories with relatively high effective area at energies at $\sim$several 10s~of keV or more may be able to detect inverse-Compton emission from the same CRe that produce radio emission. 

\end{itemize}

\end{document}